\documentclass[12pt, a4paper]{article}

\usepackage[latin1]{inputenc}
\usepackage{vmargin}
\usepackage{amsfonts}
\usepackage{amssymb}
\usepackage{graphicx}
\usepackage{amsmath}
\usepackage{amsthm}
\usepackage{setspace}
\usepackage{multirow}
\usepackage{hyperref}
\usepackage{cases}
\usepackage[round]{natbib}

\newtheorem{definition}{Definition}[section]
\newtheorem{theorem}{Theorem}[section]

\newtheorem{assumption}{Assumption}

\def\calE{{\cal E}}

\def\calQ{{\cal Q}}
\def\calS{{\cal S}}

\let\etype=\calT
\def\Insert{\calE^+}
\def\Cancel{\calE^-}
\def\Trade{\calE^t}
\def\one{\mathbf{1}}
\def\Size{\calS}
\def\Qempty{\calQ^0}
\def\Qsmall{\calQ^-}
\def\Qmed{\bar{\calQ}}
\def\Qlarge{\calQ^+}

\usepackage[bottom]{footmisc}
\setmarginsrb{2cm}{3cm}{2cm}{2cm}{0cm}{0cm}{0cm}{0.5cm}
\title{Simulating and analyzing order book data:\\ 
The queue-reactive model}

\author{Weibing Huang$^{1,2}$, Charles-Albert Lehalle$^{3}$ and Mathieu Rosenbaum$^{1}$\\$~~$\\
$^{1}$ LPMA, University Pierre et Marie Curie (Paris 6)\\
$^{2}$ Kepler-Cheuvreux\\
$^{3}$ Capital Fund Management}

\date{\today}

\begin{document}

\maketitle

\begin{abstract}
\noindent Through the analysis of a dataset of ultra high frequency order book updates, we introduce a model which accommodates the empirical properties of the full order book together with the stylized facts of lower frequency financial data. To do so, we split the time interval of interest into periods in which a well chosen reference price, typically the midprice, remains constant. Within these periods, we view the limit order book as a Markov queuing system. Indeed, we assume that the intensities of the order flows only depend on the current state of the order book. We establish the limiting behavior of this model and estimate its parameters from market data. Then, in order to design a relevant model for the whole period of interest, we use a stochastic mechanism that allows to switch from one period of constant reference price to another. Beyond enabling to reproduce accurately the behavior of market data, we show that our framework can be very useful for practitioners, notably as a market simulator or as a tool for the transaction cost analysis of complex trading algorithms.

\end{abstract}


\section{Introduction}

Electronic limit order books (LOB for short), where market participants send their buy and sell orders via a continuous-time double auction system, are nowadays the dominant mode of exchange on financial markets. Consequently, understanding the LOB dynamics has become a fundamental issue. Indeed, a deep knowledge of the LOB's behavior enables policy makers to design relevant regulations, market makers to provide liquidity at cheaper prices, and investors to save transaction costs while mounting and unwinding their positions, thus reducing the cost of capital of listed companies. Furthermore, it can also provide insights on the macroscopic features of the price which emerges from the LOB.\\

In the seminal work on zero intelligence LOB models of \citet*{smith2003statistical}, a mean-field approach is suggested in order to study the properties of the LOB. In such models, the underlying assumption is that the order flows follow independent Poisson processes. Although this hypothesis is not really compatible with empirical observations, the authors show that its simplicity allows for the derivation of many interesting formulas, some of them being testable on market data. This work has been followed by numerous developments. For example, in \citet*{cont2010stochastic}, the probabilities of various order book related events are computed in this framework, whereas stability conditions of the system are studied in \citet*{abergel2011mathematical}. We wish to extend this approach in two directions. On the one hand, we want our model to be more consistent with market data, so that we can give new insights on the dynamics of the LOB. On the other hand, we aim at providing a useful and relevant tool for market practitioners, notably in the perspective of transaction costs analysis.\\

Under the first in first out rule (which we assume in the sequel), a LOB can be considered as a high-dimensional queuing system, where orders arrive and depart randomly. We consider the three following types of orders: 

\begin{itemize}
\item Limit orders: insertion of a new order in the LOB (a buy order at a lower price than the best ask price, or a sell order at a higher price than the best bid price).
\item Cancellation orders: cancellation of an already existing order in the LOB.
\item Market orders: consumption of available liquidity (a buy or sell order at the best available price).
\end{itemize}

In practice, market participants (or their algorithms) analyze many quantities before sending a given order at a given level. One of the most important variables in this decision process is probably the distance between their target price and their ``reference market price'', typically the midprice. This reference price is linked with the order flows since it is usually determined by the LOB state. This interconnection makes the design of LOB models quite intricate. To overcome this difficulty, we split the time interval of interest into periods of constant reference price, and consider two parts in our modeling. First, we study the LOB as a Markov queuing system during the time periods when the reference price is constant. Then, we investigate the dynamics of the reference price.  Such a framework is particularly suitable for large tick assets\footnote{A large tick asset is defined as an asset whose bid-ask spread is almost always equal to one tick, see \citet*{dayri2012large}. In practice, our framework can be considered relevant for any asset whose average spread is smaller than 2.5 ticks.}, for which constant reference price periods are quite long and allow for accurate parameter estimations.\\

Two kinds of public information are available to market participants at the high frequency scale: the historical order flows and the current state of the LOB. In this paper, we are mostly interested in how the state of the LOB impacts market participants decisions. Surprisingly enough, this question has been rarely considered in the literature. Let us mention as an exception the interesting approach in \citet*{Gareche2013}, where the impact of the LOB state on the queue dynamics is analyzed through PDE type arguments. Within periods of constant reference price, we model the LOB as a continuous-time Markov jump process, and estimate its infinitesimal generator matrix under various assumptions on the information set used by market participants. From these results, we are able to analyze how market participants react towards different configurations of the LOB. Furthermore, we provide the asymptotic distributions of the LOB. The level of realism of our approaches is assessed by comparing expected features from the models with empirical ones. Thus, all our developments are illustrated on two specific examples of large tick stocks on Euronext Paris: France Telecom and Alcatel-Lucent (in appendix).\\

In the second part of the paper, we extend our framework by allowing reference price moves, so that our model also accommodates macroscopic properties of the asset (roughly summarized by the volatility). Modifications of the reference price\footnote{Note that the reference price will not be exactly the midprice, see Section \ref{sec:datadescription}.} will possibly occur provided one of the best queues is totally depleted or a new order is inserted within the spread. This model is called ``queue-reactive model". In particular, it enables us to bring to light a quantity, the ``maximal mechanical volatility", which represents the amount of price volatility generated by the generic randomness of the order flows. In practice, this parameter is typically smaller than the empirical volatility estimated from market data. The reason for this is simple: the market does not evolve like a closed physical system, where the only source of randomness would be the endogenous interactions between participants. It is also subject to external informations, such as the news, which increase the volatility of the price. Hence, it will be necessary to introduce an exogenous component within the queue-reactive model.\\

Throughout the paper, we illustrate the fact that many useful short term predictions can be computed in our framework: execution probabilities of passive orders, probability of price increase\ldots More importantly, we show that the queue-reactive model turns out to be a very relevant market simulator, notably in view of the analysis of complex trading tactics, using for example a mixture of market and limit orders.\\

The paper is organized as follows. In Section \ref{sec:lobpart1}, we consider periods when the reference price is constant. We first present a very general framework for the LOB dynamics and then introduce three specific models. The first one is a birth and death process in which the queues are assumed to be independent. In this setting, we are able to fully characterize the asymptotic behavior of the LOB. The second approach is a queuing system in which the bid and ask sides are independent, but the first two lines on each side can exhibit correlations. We show that this model can be seen as a quasi birth and death process (QBD for short) and thus admits a matrix geometric solution as its invariant distribution. In the last approach, we allow for cross dependences between bid and ask queues. An application of these models to the computation of execution probabilities is presented at the end of the same section. In Section \ref{sec:queuereactive1}, we investigate the dynamics of the reference price. In particular, we build the queue-reactive model which is a relevant LOB model for the whole time period of interest. We end this section by showing how our framework can be used for transaction costs and market impact analysis of high frequency trading strategies. A conclusion and some perspectives are given in Section \ref{sec:conclusion}. Some proofs and further empirical results are gathered in an appendix.

\section{Dynamics of the LOB in a period of constant reference price}\label{sec:lobpart1}

Within time periods when the reference price is constant, we consider three different models for the LOB. These models can be jointly introduced through the general framework we present now. 

\subsection{General Framework}\label{sec:gf}
\begin{figure}
\centering
\includegraphics[scale=0.5]{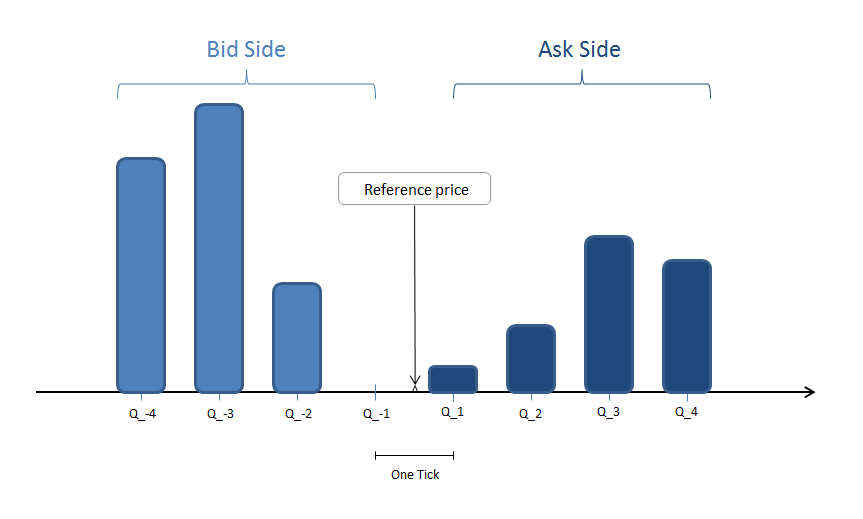}
\caption{Limit order book}\label{fig_lobexample}
\end{figure}

In the general framework, the LOB is seen as a $2K$-dimensional vector, where $K$ denotes the number of available limits on each side\footnote{Note that an empty limit can be part of the LOB in our setting.}, see Figure \ref{fig_lobexample}. The reference price $p_{ref}$ defines the center of the $2K$-dimensional vector, and divides the LOB into two parts: the bid side $[Q_{-i}:i=1,...,K]$ and the ask side $[Q_{i}:i=1,...,K]$, where $Q_{\pm i}$\footnote{To simplify our notations, we write $*_{i}$/$*_{-i}$ as $*_{\pm i}$, and $*_{-i}$/$*_{i}$ as $*_{\mp i}$.} represents the limit at the distance  $i - 0.5$ ticks to the right ($+i$) or to the left ($-i$) of $p_{ref}$. The number of orders at $Q_i$ is denoted by $q_i$. We assume that on the bid (resp. ask) side, market participants send buy (resp. sell) limit orders, cancel existing buy (resp. sell) orders and send sell (resp. buy) market orders. We consider a constant order size at each limit. However, the order sizes at the different limits are allowed to be different. In practice, these sizes can be chosen as the average event sizes observed at each limit $Q_{i}$ ($\mbox{AES}_i$ for short)\footnote{In our framework, $\mbox{AES}_i$ is a more suitable choice than ATS (Average Trade Size) that computes only the average size of market orders, see Section \ref{sec:aes} in appendix for more details.}.\\

The $2K$-dimensional process $X(t) = (q_{-K}(t),...,q_{-1}(t),q_{1}(t),...,q_{K}(t))$ is then modeled as a continuous-time Markov jump process in the countable state space $\Omega = \mathbb{N}^{2K}$, with jump size equal to one. For $q = (q_{-K},...,q_{-1},q_1,...,q_K) \in \Omega$, and $e_i = (a_{-K},...,a_{i},...,a_{K})$, where $a_{j} = 0$ for $j \neq i$ and $a_{i} = 1$, the components $\mathcal{Q}_{q,p}$ of the infinitesimal generator matrix $\mathcal{Q}$ of the process $X(t)$ are assumed to be of the following form:

\begin{eqnarray*}
\mathcal{Q}_{q,q+e_i} &=& f_i(q) \nonumber\\
\mathcal{Q}_{q,q-e_i} &=& g_i(q) \nonumber\\
\mathcal{Q}_{q,q} &=& -\sum_{q \in \Omega, p \neq q} \mathcal{Q}_{q,p} \nonumber\\
\mathcal{Q}_{q,p} &=& 0, \mbox{otherwise.}
\end{eqnarray*}

We now give a theoretical result on the ergodicity of the system under two very general assumptions. Let us denote by $P_{q,p}(t)$ the transition probability from state $q$ to state $p$ in a time $t$. Recall that a Markov process in a countable state space is said to be ergodic if there exists a probability measure $\pi$ that satisfies $\pi P = \pi$ ($\pi$ is called invariant measure) and for every $q$ and $p$:  

\begin{equation*}
\underset{t \to \infty}{\text{lim}} P_{q,p}(t) = \pi_p. 
\end{equation*}
We consider the two following assumptions.
\begin{assumption}\label{assump1}
(Negative individual drift) There exist a positive integer $C_{bound}$ and $\delta > 0$, such that for all $i$ and all $q \in \Omega$, if $q_i > C_{bound}$, $$f_i(q) - g_i(q) < -\delta.$$
\end{assumption}

\begin{assumption}\label{assump2}
(Bound on the incoming flow) There exists a positive number $H$ such that for any $q \in \Omega$, $$\sum_{i \in [-K,...,-1,1,...,K]}f_i(q) \leq  H.$$
\end{assumption}

Assumption \ref{assump1} can be interpreted as follows: the queue size of a limit tends to decrease when it becomes too large. Assumption \ref{assump2} ensures no explosion in the system: the order arrival speed stays bounded for any given state of the LOB. Under these two assumptions, we have the following ergodicity result for the $2K$-dimensional queuing system. The proof is given in appendix. 

\begin{theorem}\label{Theo1}
Under Assumptions \ref{assump1} and \ref{assump2}, the $2K$-dimensional Markov jump process $X(t)$ is ergodic.
\end{theorem}

This theorem is the basis for the asymptotic study of the LOB dynamics in the following sections.

\subsection{Data description and estimation of the reference price} \label{sec:datadescription}

\subsubsection{The database}

The data used in our empirical studies are collected from Cheuvreux's\footnote{Cheuvreux is a brokerage firm based in Paris, formerly a subsidiary of Cr\'edit Agricole Corporate Investment Bank, and now merged with Kepler Capital Market.} LOB database, from January 2010 to March 2012, on Euronext Paris. It records the LOB data (prices, volume and number of orders) up to the fifth best limit on both sides, whenever the LOB state changes. Note that we remove market data corresponding to the first and last hour of trading, as these periods have usually specific features because of the opening/closing auction phases. Two large tick European stocks, France Telecom and Alcatel-Lucent, are studied and they exhibit very similar behaviors. Some characteristics of these two stocks are given in Table 1. We have chosen the stock France Telecom as illustration example for all the developments in the sequel. The results for Alcatel-Lucent can be found in appendix. Although only stocks are considered in this paper, our method applies also to other financial assets, such as interest rates or index futures (among which large tick assets are quite numerous, see \citet*{dayri2012large}).

\begin{table}[h]\footnotesize
\begin{center}
\begin{tabular}{| c |  c|  c |  c |}
\hline
stock & average number of & average number of & average spread size\\
~&orders per day&trades per day& (in number of ticks)\\ \hline
France Telecom & 159250 & 7282 & 1.43 \\ \hline
Alcatel Lucent & 129400 & 8626 & 1.99 \\ \hline
\end{tabular}
\end{center}
\caption{Data description}
\end{table}

\subsubsection{Estimation of the reference price}

As mentioned in the introduction, the estimation of a relevant reference price $p_{ref}$ is the basis for defining the limits in the order book. Indeed, $p_{ref}$ provides the center point of the LOB and thus the positions of the $2K$ limits. In our framework, if we write $p_i$ for the price level of the limit $Q_i$, $i=-K,...,1,1,...,K$, we must have $$p_{ref} = \frac{p_{1}+p_{-1}}{2}.$$ When the observed bid-ask spread is equal to one tick, $p_{ref}$ is obviously taken as the midprice (denoted by $p_{mid}$) and both $Q_1$ and $Q_{-1}$ are non empty. When it is larger than one tick, several choices are possible for $p_{ref}$. We build $p_{ref}$ from the data the following way: when the spread is odd (in tick unit), it is still natural to use $p_{mid}$ as the LOB center: 
$$p_{ref}=p_{mid}=\frac{(p_{bestbid}+p_{bestask})}{2}.$$ When it is even, $p_{mid}$ is no longer appropriate since it is now itself a possible position for order arrivals. In such case, we use either $$p_{mid}+\frac{\mbox{tick size}}{2}\text{ or } p_{mid}-\frac{\mbox{tick size}}{2},$$ choosing the one which is the closest to the previous value of $p_{ref}$. Note that more complex methods could be used for the estimation of $p_{ref}$, see for example \citet*{delattre2013estimating}.

\subsection{Model I: Collection of independent queues}

We now give a first simple LOB model around a fixed reference price.

\subsubsection{Description of the model}

In this model, we assume independence between the flows arriving at different limits in the LOB. Three types of orders are considered: limit orders, cancellations and market orders. We suppose that the intensities of these point processes at different limits are only functions of the target queue size (that is the available volume at the considered limit $Q_i$). Furthermore, at a given limit, conditional on the LOB state, the arrival processes of the three types of orders are taken independent. The values of these intensities are denoted by $\lambda^L_i(n)$ (limit orders), $\lambda^C_i(n)$ (cancellations) and $\lambda^M_i(n)$ (market orders) when $q_i=n$. Moreover, the intensity functions at $Q_i$ and $Q_{-i}$ are chosen identical, considering the symmetry property of the LOB. We then have $\lambda_i^L(n) = \lambda_{-i}^L(n), \lambda_i^C(n) = \lambda_{-i}^C(n), \lambda_i^M(n) = \lambda_{-i}^M(n)$, and

\begin{eqnarray*}
f_i(q) &=& \lambda_i^L(q_i) \nonumber\\
g_i(q) &=& \lambda_i^C(q_i) + \lambda_i^M(q_i).  \label{formulamodel1}
\end{eqnarray*}

In this model, market orders sent to $Q_i$ consume directly the volume available at $Q_i$. Therefore, we can have a market order at the second limit while the first limit is not empty. However, for large tick assets, this assumption is reasonable as their market order flow is almost fully concentrated on first limits ($Q_{\pm 1}$) and the estimated intensities of this flow at ($Q_{\pm i}$), $i\neq 1$ are very small. Under these assumptions, the LOB becomes a collection of $2K$ independent queues, each of them being a birth and death process. 

\subsubsection{Empirical study: Collection of independent queues}\label{sec:intest}

In Model I, the intensities of the different queues can be estimated separately. The value of $K$ is set to 3, as our numerical experiments show that for the considered stocks, both the dynamics and empirical distributions at $Q_{\pm i}, i = 4,5$ are quite similar to that at $Q_{\pm 3}$. This value of $K$ will also apply to other experiments in the paper.\\ 

The estimation method goes as follows. We define an ``event''  $\omega$  as any modification of the queue size. For queue $Q_i$, we record the waiting time $\Delta t_i(\omega)$ (in number of seconds) between the event $\omega$ and the preceding event at $Q_i$, the type of the event $\etype_i(\omega)$ and the queue size $q_i(\omega)$ before the event. The queue size is then approximated by the smallest integer that is larger than or equal to the volume available at the queue, divided by the stock's average event size $\mbox{AES}_i$ at the corresponding queue. We set the ``type'' of the event $\omega$ the following way:  
\begin{itemize}
\item $\etype_i(\omega)\in\Insert$ for limit order insertion at $Q_i$,
\item $\etype_i(\omega)\in\Cancel$ for limit order cancellation at $Q_i$, 
\item $\etype_i(\omega)\in\Trade$ for market order at $Q_i$. 
\end{itemize}
When the reference price changes, we restart the recording process. Once we have collected $(\Delta t_i(\omega), \etype_i(\omega), q_i(\omega))$ from historical data, it is easy to estimate $\lambda_{i}^L(n)$, $\lambda_{i}^C(n)$ and $\lambda_i^M(n)$ by the maximum likelihood method:

\begin{eqnarray}
\hat{\Lambda}_i(n) &=& \big({\mbox{mean}(\Delta t_i(\omega) | q_i(\omega) = n)}\big)^{-1} \nonumber \\
\hat{\lambda}_i^L(n) &=& \hat{\Lambda}_i(n) \frac{\# \{\etype_i(\omega) \in\Insert, q_i(\omega) = n\}}{\# \{q_i(\omega) = n\}} \nonumber\\
\hat{\lambda}_i^C(n) &=& \hat{\Lambda}_i(n) \frac{\# \{\etype_i(\omega) \in\Cancel, q_i(\omega) = n\}}{\# \{q_i(\omega) = n\}} \nonumber\\
\hat{\lambda}_i^M(n) &=& \hat{\Lambda}_i(n) \frac{\# \{\etype_i(\omega) \in\Trade, q_i(\omega) = n\}}{\# \{q_i(\omega) = n\}}, \nonumber
\end{eqnarray}
where ``mean'' denotes the empirical mean and $\#A$ the cardinality of the set $A$.\\

\begin{figure}
\centering
\includegraphics[scale=0.65,angle =-90]{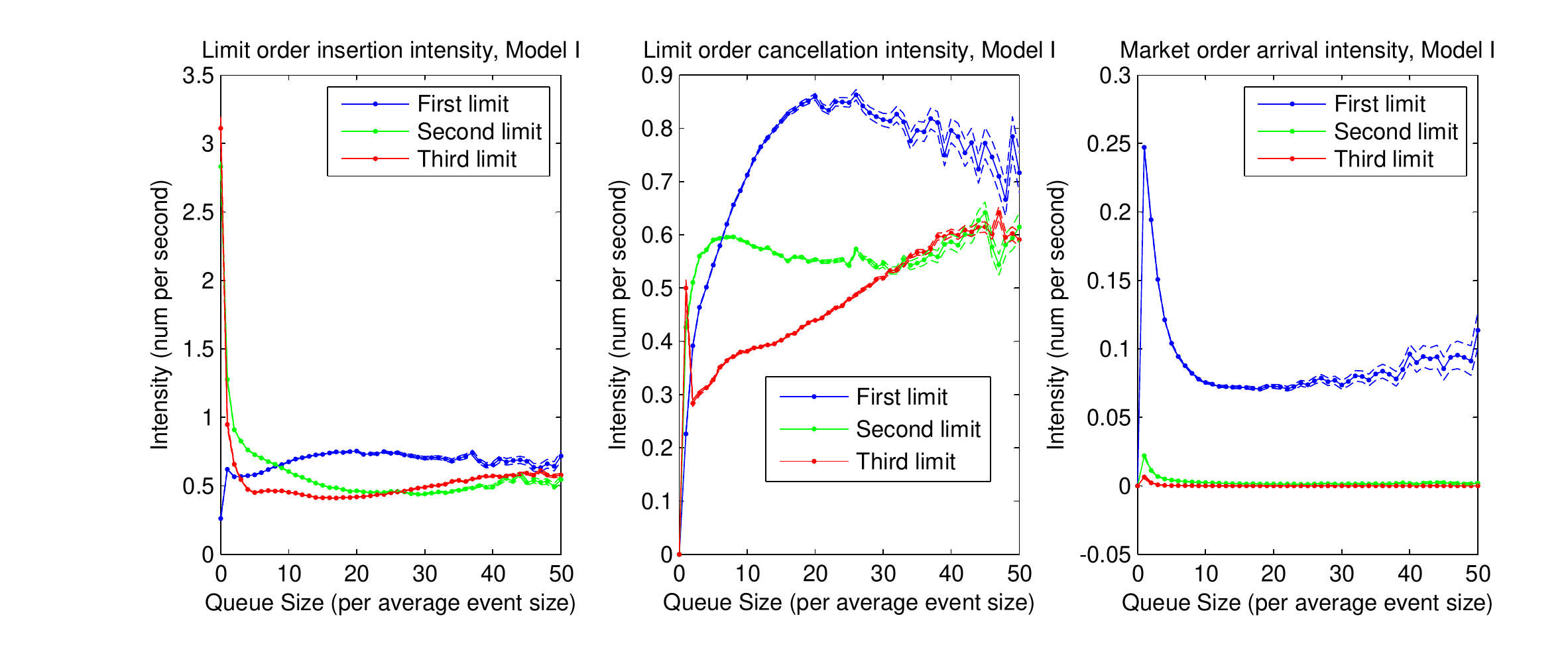}
\caption{Intensities at $Q_{\pm i}$, $i=1,~2,~3$, France Telecom}\label{empfig1}
\end{figure}

In Figure \ref{empfig1}, we present the estimated intensities. Data at $Q_i$ and $Q_{-i}$ are aggregated together (simply by combining the two collected samples) and confidence intervals (dotted lines) are computed using central limit approximations detailed in appendix. We now comment the obtained graphs.

\paragraph{Behaviors under the independence assumption}  \label{sec:model1}

\begin{itemize}
\item Limit order insertion: 
\begin{itemize}
\item[$Q_{\pm 1}$:] The intensity of the limit order insertion process is approximately a constant function of the queue size, with a significantly smaller value at 0. Note that inserting a limit order in an empty queue creates a new best limit and the market participant placing this order is the only one standing at this price level. Such action is often risky. Indeed, when the spread is different from one tick, one is quite uncertain about the position of the so-called ``efficient" or ``fair" price, see for example \citet*{delattre2013estimating} for discussions on this notion. This smaller value can also be due to temporary realizations of the structural relation between the bid-ask spread and the volatility: if the spread is large because the inventory risk of market makers is high, the probability that anyone inserts a limit order in the spread is likely to be low, see among others \citet*{madhavan1997security}, \citet*{avellaneda2008high}, \citet*{wyart2008relation} and \citet*{dayri2012large} for more details about market making and the relation between spread, volatility and inventory risk.
\item[$Q_{\pm 2}$:] The intensity is now approximately a decreasing function of the queue size. This interesting result probably reveals a quite common strategy used in practice: posting orders at the second limit when the corresponding queue size is small to seize priority. More details on this strategy are given in Section \ref{sec:intensityq2}. 
\item[$Q_{\pm 3}$:] The intensity function shows similar properties to that at the second limit.

\end{itemize}
\item Limit order cancellation: 
\begin{itemize} 
\item[$Q_{\pm 1}$:] The rate of order cancellation is an increasing concave function for $q_{\pm 1}$ between $0$ and $25$, and becomes flat/slightly decreasing for larger values. This result is in contrast to the classical way to model this flow, where one often considers a linearly increasing cancellation rate, see for example \citet*{cont2010stochastic}. On this first in first out market, the priority value, that is the advantage of a limit order compared with another limit order standing at the rear of the same queue, can be one of the reasons for this behavior. Indeed, the priority value is an increasing function of the queue size and orders having a high priority value are less likely to be canceled. 
\item[$Q_{\pm 2}$:] The rate of order cancellation attains more rapidly its asymptotic value, which is lower than for $Q_{\pm 1}$. Compared to the first limit case, market participants at the second limit have even stronger intention not to cancel their orders when the queue size increases. This is probably due to the fact that these orders are less exposed to short term market trends than those posted at $Q_{\pm 1}$ (since they are covered by the volume standing at $Q_{\pm 1}$ and their price level is farther away from the reference price).
\item[$Q_{\pm 3}$:] The priority value is smaller at the third limit since it takes longer time for $Q_{\pm 3}$ to become the best quote if it does. The rate of order cancellation increases almost linearly for queue sizes larger than 3 $\mbox{AES}_3$. We also find a quite large cancellation rate when the queue size is equal to one, which shows that market participants cancel their orders more quickly when they find themselves alone in the queue. 
\end{itemize}

\item Market orders: 
\begin{itemize} 
\item[$Q_{\pm 1}$:] The rate decreases exponentially with the available volume at $Q_{\pm 1}$. This phenomena is easily explained by market participants ``rushing for liquidity'' when liquidity is rare, and ``waiting for better price'' when liquidity is abundant.  
\item[$Q_{\pm 2}$:] In practice, market orders can arrive at $Q_{\pm 2}$ only if $Q_{\pm 1} = 0$ (that is when $Q_{\pm 2}$ is the best offer queue). The shape of the intensity is very similar to the one obtained in the case of $Q_{\pm 1}$. The values are of course much smaller. 
\item[$Q_{\pm 3}$:] In some rare cases, one can still find some market orders arriving at $Q_{\pm 3}$ (market orders occurring when the spread is large). The intensity function remains exponentially decreasing.
\end{itemize}
\end{itemize}

\subsubsection{Asymptotic behavior under Model I}

The invariant distribution of the LOB can be computed explicitly in Model I. We denote by $\pi_i$ the stationary distribution of the limit $Q_i$, and define the arrival/departure ratio vector $\rho_i$ by $$\rho_i(n) = \frac{\lambda_i^L(n)}{(\lambda_i^C(n+1) + \lambda_i^M(n+1))}.$$ Then the following result for the invariant distribution is easily obtained, see for example \citet*{gross2013fundamentals}: 

\begin{eqnarray*}
\pi_i(n) &=& \pi_i(0)\prod_{j=1}^n\rho_i(j-1)  \nonumber \\
\pi_i(0) &=& \big({1+\sum_{n=1}^\infty \prod_{j=1}^n\rho_i(j-1)}\big)^{-1}.
\end{eqnarray*} \label{model1_inv}

Hence the long term behavior of the LOB is completely determined by $\rho$. This implies that two assets can have very different flow dynamics, but still the same invariant distribution provided that their arrival/departure ratios are the same. \\

\begin{figure}
\centering
\includegraphics[scale=0.65,angle=-90]{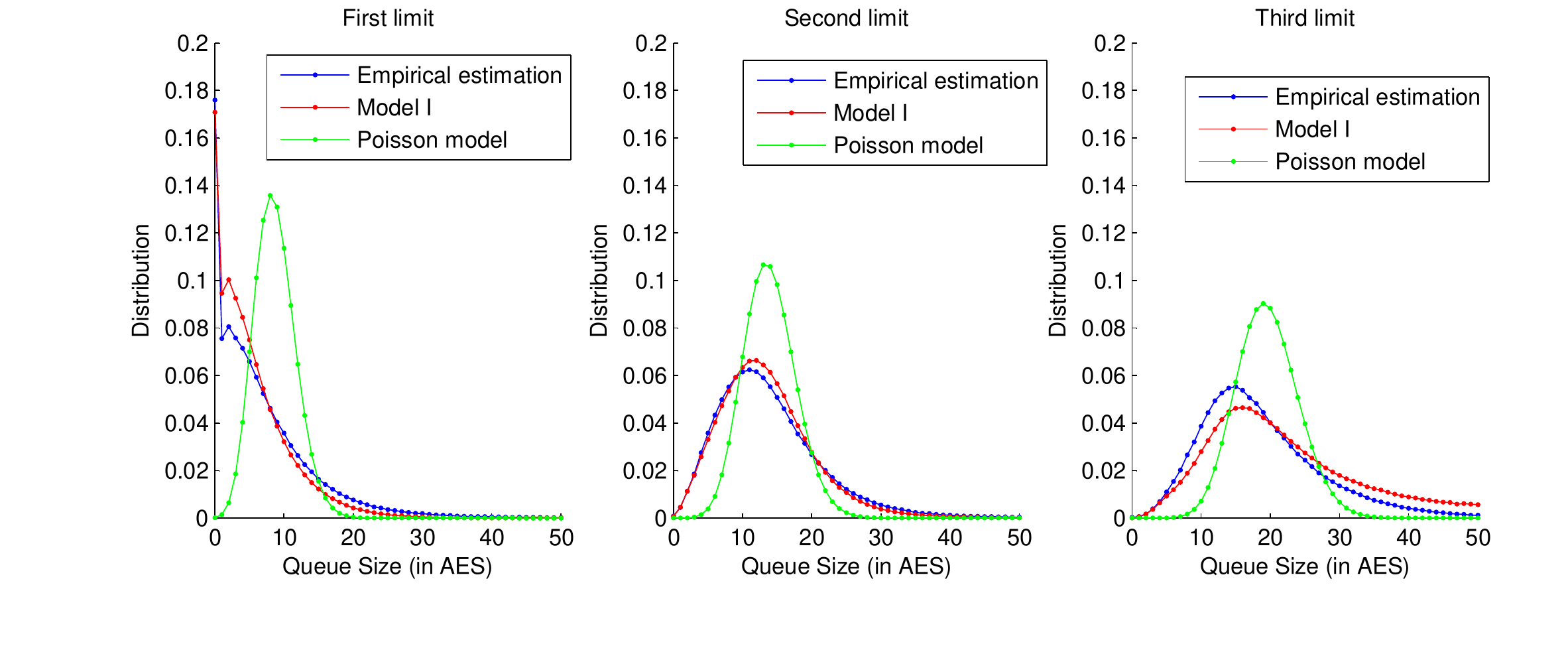}
\caption{Model I, invariant distributions of $q_{\pm 1},q_{\pm 2},q_{\pm 3}$, France Telecom}
\label{univariate_model}
\end{figure}

We now compare the asymptotic results of the model with the empirical distributions observed at $Q_{\pm 1}$, $Q_{\pm 2}$ and $Q_{\pm 3}$. To compute these empirical laws, we use a sampling frequency of 30 seconds (every 30 seconds, we look at the LOB and record its state)\footnote{Other sampling frequencies have also been tested and the estimated distributions are found to be very similar. These sampled data will also be used to estimate the joint distributions of the LOB limits in Model II$^a$ and II$^b$.}. The results are gathered in Figure \ref{univariate_model}, as well as the invariant distributions from a Poisson model (constant limit/market order arrival rate, linear cancellation rate, parameters estimated from the same dataset).\\

One can see that the invariant distributions approximate very well the empirical distributions of the LOB. This shows that in order to explain the shape of the LOB, such mean-field type approach, where the LOB profile arises from interactions between the average behaviors of market participants, can be very relevant. 

\subsection{Model II: Dependent case}\label{sec:twosetsofdependentqueues}

We now present some extensions of Model I. We assume here that buy/sell market orders consume volume at the best quote limits, defined as the first non empty ask/bid queue. Thus, we consider a buy market order process with intensity $\lambda_{buy}^M$ and a sell market order process with intensity $\lambda_{sell}^M$. The limit order, cancellation, and market order arrival processes are assumed to be independent conditional on the LOB state. So we can write $f_i(q)$ and $g_i(q)$ in the following form:

\begin{eqnarray*}
f_i(q) &=& \lambda^L_i(q) \nonumber\\
g_i(q) &=& \lambda^C_i(q) + \lambda^M_{buy}(q)\one_{bestask(q)=i}, \mbox{if $i>0$} \nonumber\\
g_i(q) &=& \lambda^C_i(q) + \lambda^M_{sell}(q)\one_{bestbid(q)=i}, \mbox{if $i<0$}. 
\end{eqnarray*}

As for Model I, we consider some bid-ask symmetry, that is, for $q =[q_{-3},q_{-2},q_{-1},q_1,q_2,q_3]$, $q' = [q_{3},q_{2},q_{1},q_{-1},q_{-2},q_{-3}]$ and $i=1,2,3$, $\lambda^L_i(q) = \lambda^L_{-i}(q')$, $\lambda^M_i(q) = \lambda^M_{-i}(q')$ and $\lambda^M_{buy}(q) = \lambda^M_{sell}(q')$ .

\subsubsection{Model II$^a$:  Two sets of dependent queues}

Institutional traders and brokers tend to place most of their limit orders at best limits, while many market makers, arbitragers and other high frequency traders stand also in queues beyond these best limits. This suggests for example that the dynamics at $Q_{\pm 2}$ may not only depend on $q_{\pm 2}$, but also on whether or not $Q_{\pm 1}$ is empty. We thus propose to use the following intensity functions for the queue $Q_{\pm 2}$: in this model, $\lambda^L_{\pm 2}$ and $\lambda^C_{\pm 2}$ are functions of $q_{\pm 2}$ and $\one_{q_{\pm 1}>0}$. Intensities at $Q_i,i\neq \pm 2$ remain functions of $q_i$ only. For large tick assets, the probability that $Q_{\pm i}, i \geq 3$ is the best limit is negligible. It is thus reasonable to also assume that market orders are only sent to $Q_{\pm 1}$ and $Q_{\pm 2}$. This enables us to keep the independence property between $Q_{\pm 3}$ and $(Q_{\pm 1}, Q_{\pm 2})$. When $q_{\pm 1} > 0$, the market order intensity $\lambda^M_{buy/sell}$ is assumed to be a function of $q_{\pm 1}$; when $q_{\pm 1} = 0$, it is a function of $q_{\pm 2}$ only. 

\subsubsection{Model II$^a$: Empirical study}\label{sec:intensityq2}

In this empirical study, our goal is to understand how market participants make trading decisions at $Q_{\pm 2}$ in two different situations: $q_{\pm 1} = 0$ and $q_{\pm 1} > 0$.  Since we are now studying a two-dimensional problem, the data recording process is slightly different. In particular, for ($Q_1,Q_2$), it goes as follows: we record the waiting times $\Delta t_i(\omega)$ between events that happen at $Q_1$ or $Q_2$, the type of event $\etype(\omega)$ and the two queue sizes ($q_1(\omega),q_2(\omega)$) before the event. The maximum likelihood method is again used to estimate the intensity functions $\lambda_i^L$, $\lambda_i^C$, $\lambda_i^M$ for $i = 1,2$. For $i=1$ and $i=3$, as the dynamics at $Q_{\pm i}$ only depend on the queue size at $Q_{\pm i}$, the estimated values of $\lambda_1^L,\lambda_1^C$ and $\lambda_1^M$ are very close to those obtained in Model I and are not shown here. The estimated intensity functions at $Q_{\pm 2}$ are given in Figure \ref{intMII}. Some comments are in order:

\begin{figure}
\centering
\includegraphics[scale=0.65,angle =-90]{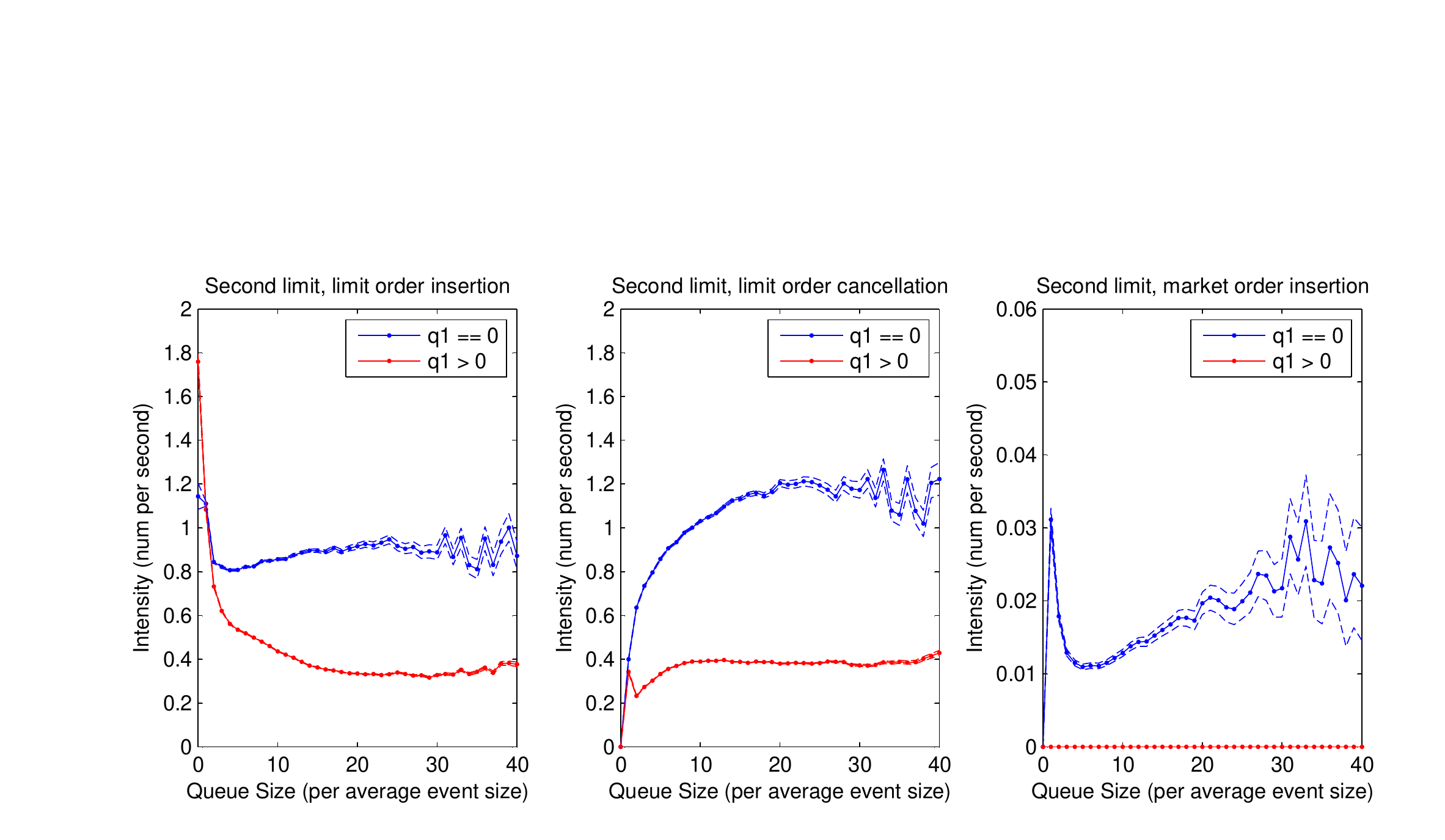}
\caption{Intensities at $Q_{2}$ as functions of $1_{q_{1} > 0}$ and $q_{2}$, France Telecom}\label{intMII} \label{fig2}
\end{figure}

\begin{itemize}
\item Limit order insertion: Both curves are decreasing functions of the queue size. In the first case ($q_{\pm 1} = 0$), the limit order insertion intensity reaches very rapidly its asymptotic value. The relatively high value observed for $q_{\pm 2} = 0$ is probably due to the fact that for large tick assets, market makers rarely allow for spreads larger than 3 ticks. In the second case ($q_{\pm 1} > 0$), the intensity continues to go down to a much lower value. This is likely to be related to the arbitrage strategy introduced in Section \ref{sec:model1}: post passive orders at a non-best limit when its size is small, wait for this limit to eventually become the best limit and then gain the profit from having the priority value. For example, when the considered limit becomes the best one, one can decide to stay in the queue if its size is large enough to cover the risk of short term market trend, or to cancel the orders if the queue size is too small. 
\item Limit order cancellation: The cancellation rate is higher when $q_{\pm 1} = 0$. This can be related to the concentration of the trading activity at best limits. When $q_{\pm 1}>0$, the cancellation rate is quite large when $q_{\pm 2} = 1$, as it is the case at $Q_{\pm 3}$ (see Section \ref{sec:model1}). 
\item Market orders: No market order can arrive at $Q_{\pm 2}$ when there are still limit orders at $Q_{\pm 1}$ (cross limits large market orders that consume several limits are treated as several market orders that arrive sequentially at those limits within a very short time period). The market order arrival rate when $Q_{\pm 2}$ is the best limit is not very different from that at $Q_{\pm 1}$, but shows a rather unexpected increasing trend when the queue size becomes larger than 5 $\mbox{AES}_2$. 
\end{itemize}

\subsubsection{Model II$^a$: Asymptotic behavior}\label{MIIQBD}

Model II$^a$ belongs to a special class of Markov processes, called quasi birth and death processes (QBD). Their asymptotic behavior can be studied by the matrix geometric method. Definitions of QBD processes and explanations about the matrix geometric method can be found in appendix. In Figure \ref{fig5}, we show the theoretical joint distribution of $(q_1,q_2)$ for the stock France Telecom and compare it with the joint distribution estimated from empirical data. Here also, we see that the theoretical results provide a very satisfying approximation. 

\begin{figure}
\centering
\includegraphics[scale=0.65,angle=-90]{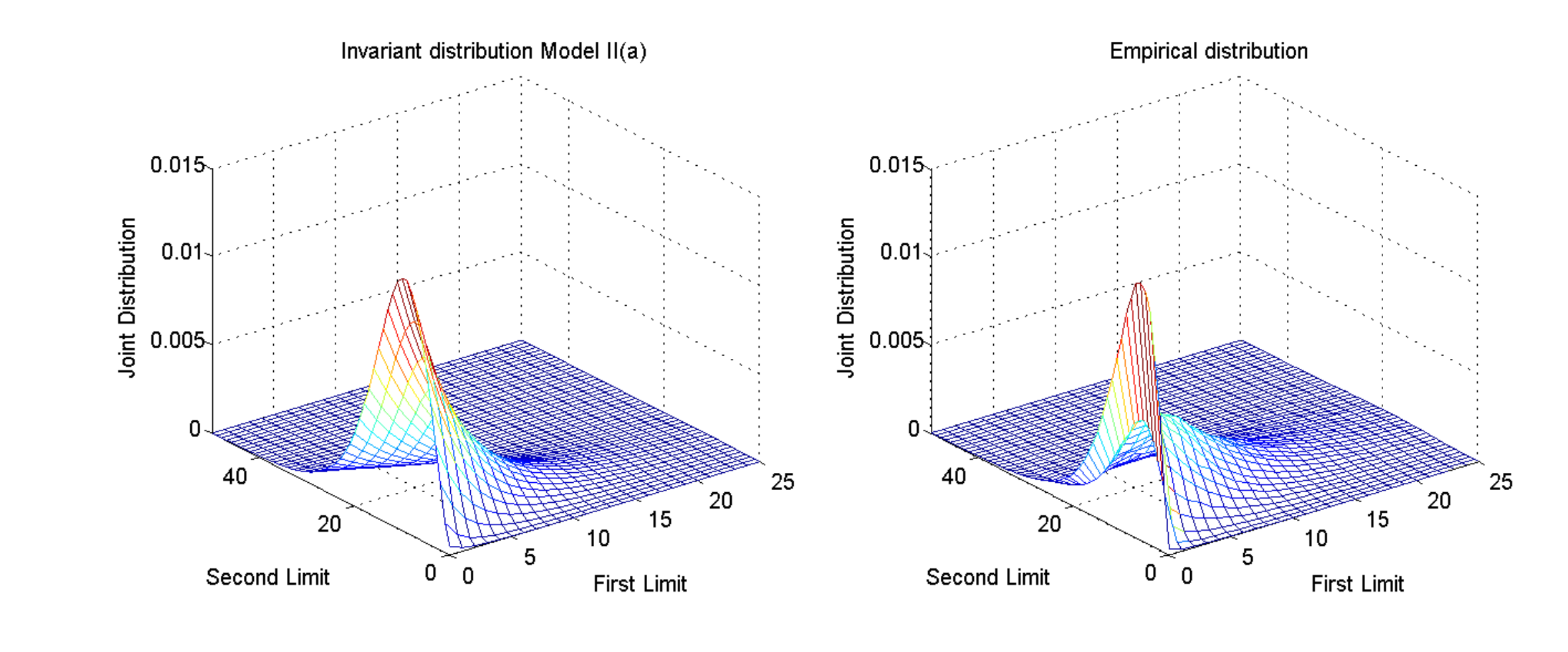}
\caption{Model II$^a$: joint distribution of $q_1,q_2$, France Telecom}\label{fig5}
\end{figure}

\subsubsection{Model II$^b$: Modeling bid-ask dependences}\label{sec:model3}

We now study the interactions between the bid queues and the ask queues. Let $\Qempty$, $\Qsmall$, $\Qmed$, $\Qlarge$ be four marks which 
represent in the following ranges of values for the queue sizes. Let $m$ and $l$ be two integers. We define the function $\Size_{m,l}(x)$:

\begin{eqnarray*}
\Size_{m,l}(x) &=& \Qempty \mbox{ if } x = 0 \nonumber \\
\Size_{m,l}(x) &=& \Qsmall \mbox{ if } 0 < x \leq m \nonumber \\
\Size_{m,l}(x) &=& \Qmed \mbox{ if } m < x \leq l \nonumber \\
\Size_{m,l}(x) &=& \Qlarge \mbox{ if } x > l.
\end{eqnarray*}

This function associates to a queue size $x$ four possible ranges: empty: $x=0$, small: $x\in(0,m]$, usual: $x\in(m,l]$ and large: $x\in(l,+\infty)$. We set $m$ as the 33\% lower quantile and $l$ as the 33\% upper quantile of $q_{\pm 1}$ (conditional on positive values). In this model, market participants at $Q_{\pm 1}$ adjust their behavior not only according to the target queue size, but also to the size of the opposite queue. The rates $\lambda^L_{\pm 1}$ and $\lambda^C_{\pm 1}$ are therefore modeled as functions of $q_{\pm 1}$ and $\Size_{m,l}(q_{\mp 1})$. As in Model II$^a$, we suppose that market orders consume volume at the best limits and are only sent to $Q_{\pm 1}$ and $Q_{\pm 2}$. When $q_{\pm 1} > 0$, the market order intensity $\lambda^M_{buy/sell}$ is assumed to be a function of $q_{\pm 1}$ and $\Size_{m,l}(q_{\mp 1})$. Regime switching at $Q_{\pm 2}$ is kept in this model: $\lambda^L_{\pm 2}$, $\lambda^C_{\pm 2}$ are assumed to be functions of $\one_{q_{\pm 1}>0}$ and $q_{\pm 2}$, and when $q_{\pm 1} = 0$, the market order intensity $\lambda^M_{buy/sell}$ is modeled as a function of $q_{\pm 2}$.\\

Under these assumptions, the $2K$-dimensional problem is reduced to the study of the 4-dimensional continuous-time Markov jump process ($Q_{-2},Q_{-1},Q_1,Q_2$). One important feature of this model is that the queues $Q_{\pm 2}$ have no influence on the dynamics at $Q_{\pm 1}$. Therefore, we only need to study the 3-dimensional process $(Q_{-1},Q_1,Q_2)$ (or even the 2-dimensional process $(Q_{-1},Q_1)$ if one is only interested in the dynamics at $Q_{\pm 1}$. Remark also that other choices for the specification of the intensity functions at $Q_{\pm 1}$ are possible. For example, one can consider them as functions of the first level bid/ask imbalance, defined as $\frac{q_1-q_{-1}}{q_1 + q_{-1}}$, or simply as functions of the spread size.

\subsubsection{Model II$^b$: Empirical study}

We focus here on the estimation of the intensity functions at $Q_{\pm 1}$. We consider the departure flow intensities $\lambda^C_{\pm 1}(q_{\pm 1},\Size_{m,l}(q_{\mp 1}))$ and $\lambda^M_{buy/sell}(q_{\pm 1},\Size_{m,l}(q_{\mp 1}))$, and the arrival flow intensities $\lambda^L_{\pm 1}(q_{\pm 1},\Size_{m,l}(q_{\mp 1}))$. Using again the symmetry property of the LOB, we take $\lambda^L_1(x,y) = \lambda^L_{-1}(x,y)$, $\lambda^C_1(x,y) = \lambda^C_{-1}(x,y)$ and $\lambda^M_{sell}(x,y) = \lambda^M_{buy}(x,y)$. We record the waiting times $\Delta t(\omega)$ between events that happen at $Q_1$ or $Q_{-1}$, the types of event $\etype(\omega)$ and the two queue sizes ($q_1(\omega),q_{-1}(\omega)$) before the event. Then we estimate these intensity functions using the maximum likelihood method. The results are shown in Figure \ref{fig6} ($m = 4 \mbox{ AES}_1$, $l = 9 \mbox{ AES}_1$)\footnote{Note that the computation of the confidence intervals becomes more intricate for this model and the results presented are slightly approximate ones, see details in appendix.}. Some remarks are in order:

\begin{figure}
\centering
\includegraphics[scale=0.65,angle =-90]{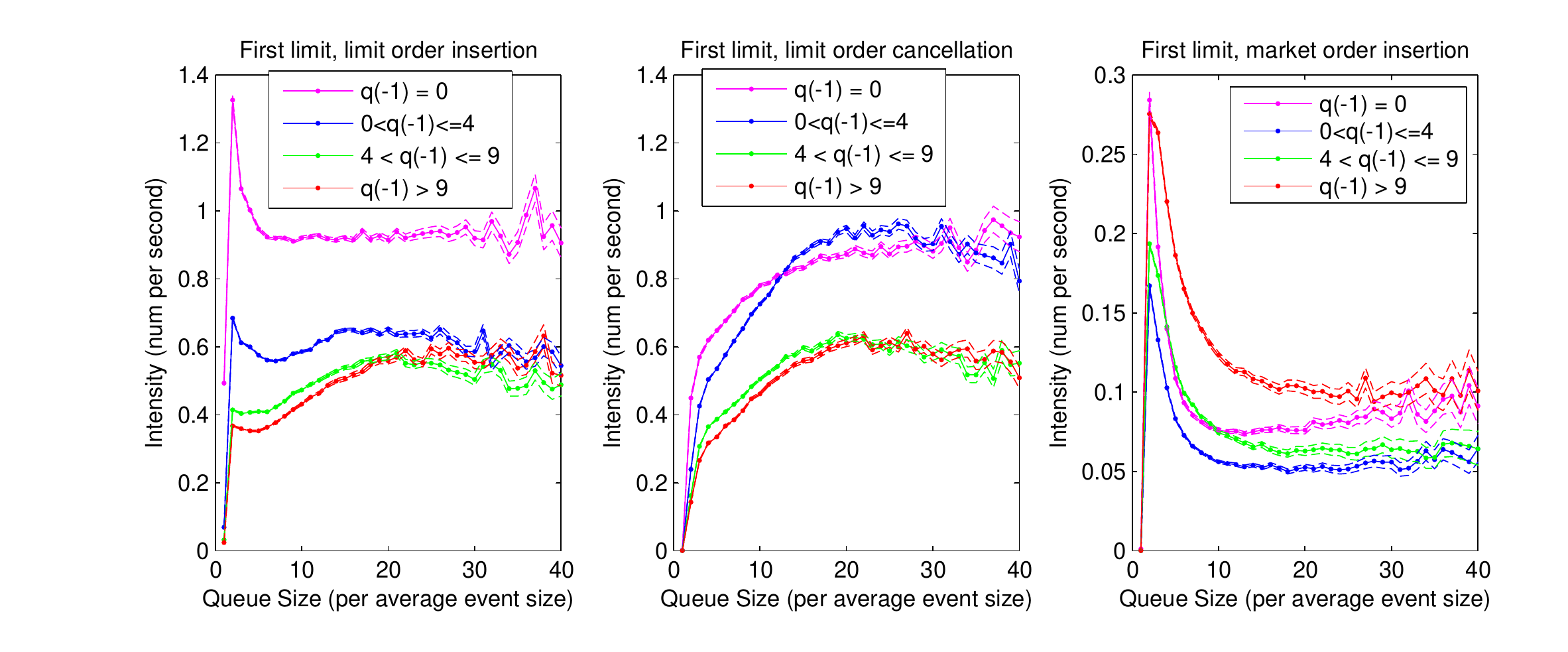}
\caption{Intensities at $Q_1$ as functions of $\Size_{m,l}(q_{-1})$ and $q_1$, France Telecom}\label{fig6}
\end{figure}

\begin{itemize}

\item Limit order insertion: The limit order insertion rate is a decreasing function of the opposite queue size. In particular, we see that when the opposite queue is empty (pink curve), it is significantly larger. Indeed, in that case, the ``efficient" price is likely to be closer to the opposite side. Therefore limit orders at the non empty first limit are likely to be profitable.
\item Limit order cancellation: The cancellation rates for different ranges of $Q_{-1}$ are similar in their forms but have different asymptotic values. This rate is not surprisingly a decreasing function of the liquidity level on the opposite side. Indeed, when this level becomes low, many market participants cancel
their limit orders and send market orders since the market is likely to move in an unfavorable direction. 
\item Market orders: We see that when the liquidity available on the opposite side is abundant, more market orders are sent. Indeed, in that case, transactions at the target queue are relatively cheap as its price level is temporarily closer to the efficient price. In the special situation $q_{-1} = 0$, the price level at $Q_1$ can seem relatively attractive since it is much closer to the reference price than the opposite best price, which is in that case 2 ticks away from it. This explains why the market order intensity is larger when the opposite queue is empty than when its size is small. 
\end{itemize}

\subsubsection{Model II$^b$: Asymptotic behavior}

\begin{figure}\label{MC}
\centering
\includegraphics[scale=0.65,angle=-90]{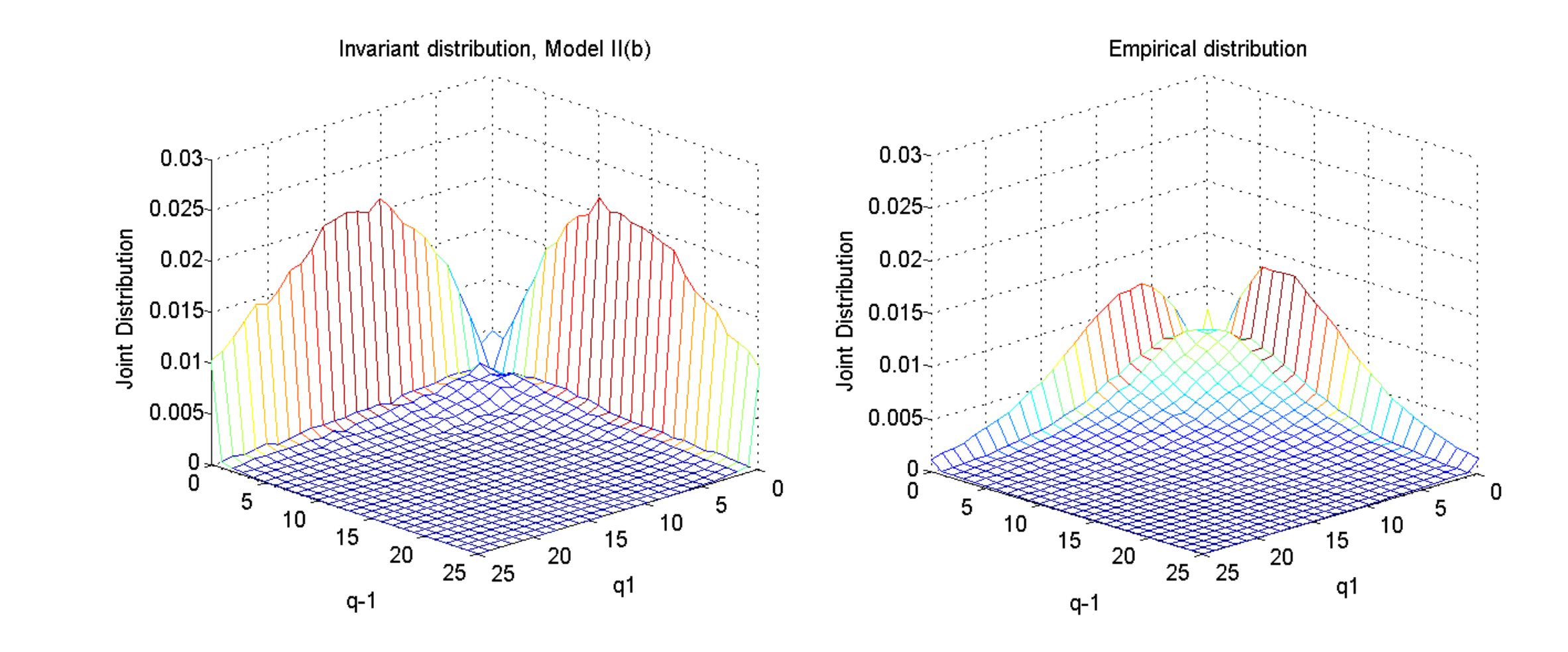}
\caption{Model II$^b$: joint distribution of $q_{-1},q_1$, France Telecom}\label{figmodel32}
\end{figure}

\begin{figure}
\centering
\includegraphics[scale=0.65,angle =-90]{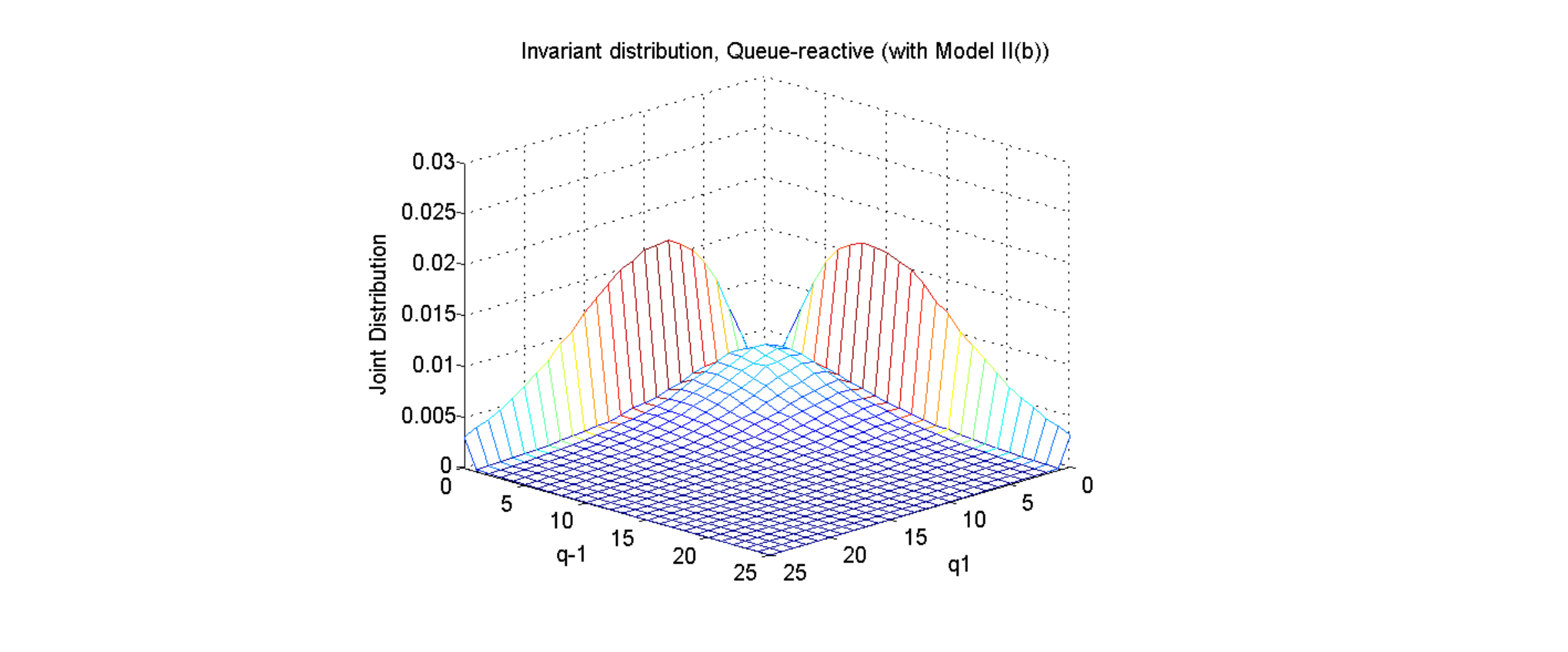}
\caption{Model III: joint distribution of $q_{-1},q_1$, France Telecom}\label{MCIIbIII6}
\end{figure}

Monte-Carlo simulations are used to obtain the theoretical invariant distribution of the LOB in Model II$^b$. The theoretical and empirical joint distributions of $Q_{-1}$ and $Q_1$ are shown in Figure \ref{figmodel32}. The difference between the two graphs comes from the relatively high probabilities of states of the form $(x,y)$ with $x$ and $y$ both small in empirical data, which are somehow replaced by states of the form $(x,0)$ or $(0,y)$ in the model. Indeed, in practice, a situation where one of the first queue is empty is not likely to remain long since it often leads to a reference price change. This effect is not taken into account in Model II$^b$ where the reference price is constant, but will be investigated in Model III in Section \ref{sec:queuereactive}. We anticipate here by giving in Figure \ref{MCIIbIII6} the joint distribution obtained when suitable moves of the reference price are added within the framework of Model II$^b$ (following the approach of Model III in Section \ref{sec:queuereactive}). We now find that the simulated density becomes very close to the empirical one.

\subsection{Example of application: Probability of execution}\label{sec:probexec}

The preceding models can be used to compute short term predictions about several important LOB related quantities. One relevant example is the probability of executing an order before the midprice moves. Suppose that at time $t=0$, both $Q_1$ and $Q_{-1}$ are not empty. Then a trader (called A) submits a buy limit order at $Q_{-1}$ of size $n_0$ and waits in the queue until either the order is executed or the opposite queue $Q_1$ is totally depleted. The probability of execution can be computed in all of the three preceding models, using Monte-Carlo simulations. \\ 

There are two types of orders at $Q_{-1}$: orders placed before $t=0$, thus having higher priority compared with the order of trader A, and orders placed after $t=0$, having lower priority. When a market order arrives at $Q_{-1}$, the limit order with the highest priority is executed. Hence trader A's order starts being executed only when all orders placed at $Q_{-1}$ before $t=0$ have been either canceled or executed. When a cancellation event happens at $Q_{-1}$, the precise order being canceled is not clearly defined in our models. So, we need to make two additional assumptions for the cancellation process.

\begin{assumption}\label{assump4}
When a cancellation event occurs at $Q_{-1}$, orders at $Q_{-1}$ have the same probability of being canceled (except for the limit order submitted by trader A, which is never canceled).
\end{assumption}

\begin{assumption}\label{assump5}
The cancellation intensity at $Q_{-1}$ is supposed to be equal to $\lambda_1^C(q_{-1})\frac{q_{-1}-n_0}{q_{-1}}$ instead of $\lambda_1^C(q_{-1})$, since the order placed by trader A is never canceled.
\end{assumption}

\begin{figure}
\centering
\includegraphics[scale=0.65,angle=-90]{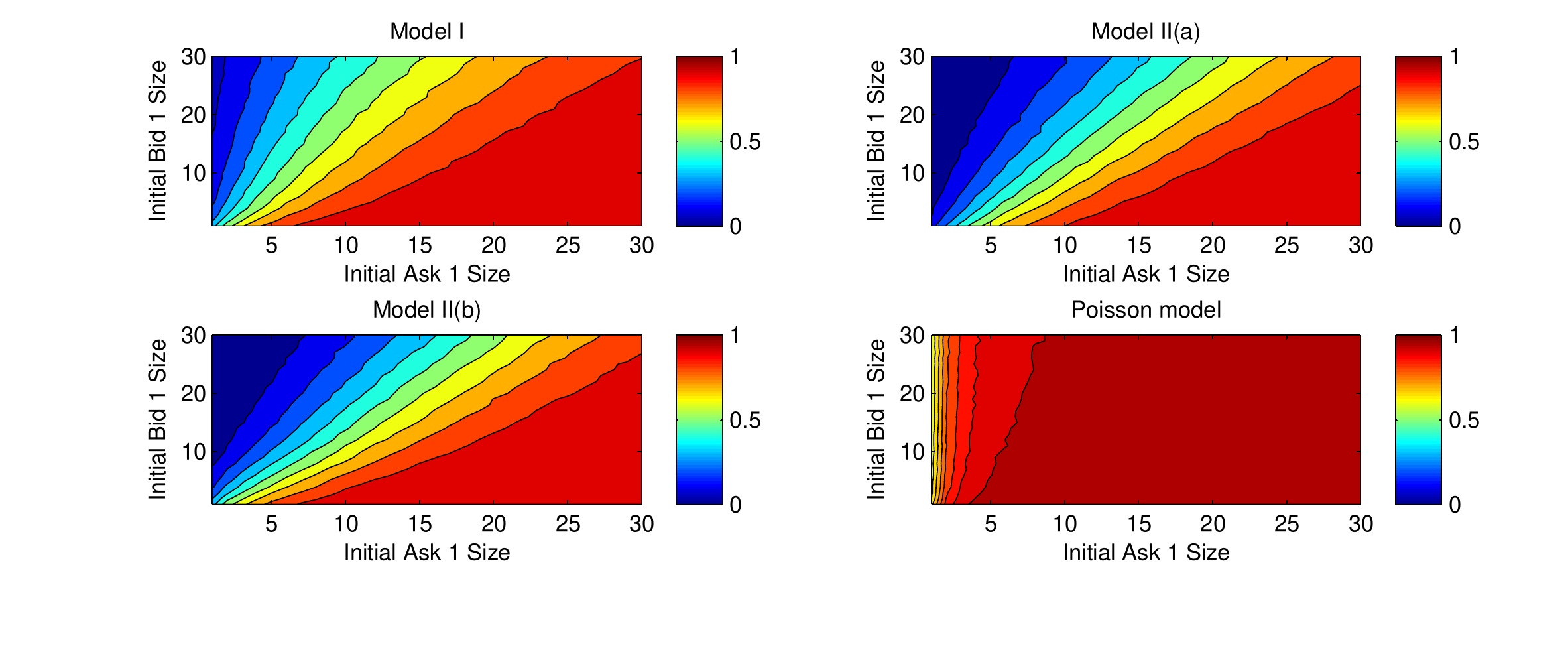}
\caption{Execution probability of a buying order placed at $Q_{-1}$ at $t = 0$, France Telecom} \label{fig17}
\end{figure}

Orders with lower priority are actually more likely to be canceled, see \citet*{Gareche2013}. However, in order to investigate precisely this feature, we would need more detailed market data keeping records of the identifiers of the submitted and canceled orders. As a result, execution probabilities might be slightly overestimated using Assumptions \ref{assump4} and \ref{assump5}. Simulation results (for $n_0=1$) are shown in Figure \ref{fig17}, together with the predictions associated to a Poisson model that assumes a linearly increasing cancellation rate. We see that our three models give fairly similar execution probabilities, while the Poisson model clearly overestimates them. 

\section{The queue-reactive model: a time consistent model with stochastic LOB and dynamic reference price}\label{sec:queuereactive1}

We now wish to obtain a model which is relevant on the whole period of interest and provides useful applications. 

\subsection{Model III: The queue-reactive model}\label{sec:queuereactive}

\subsubsection{Building the model}

Let $\delta$ denote the tick value. We assume here that $p_{ref}$ changes with some probability $\theta$ when some event modifies the midprice $p_{mid}$. More precisely, when $p_{mid}$ increases/decreases\footnote{Note that in this model, $p_{ref}$ does not necessarily match its estimated value using the method introduced in Section \ref{sec:datadescription}. However, for large tick assets, the difference is negligible.}, $p_{ref}$ increases/decreases by $\delta$ with probability $\theta$, provided $q_{\pm 1} = 0$ at that moment. Hence changes of $p_{ref}$ are possibly triggered by one of the three following events:

\begin{itemize}
\item The insertion of a buy limit order within the bid-ask spread while $Q_{1}$ is empty at the moment of this insertion, or the insertion of a sell limit order within the bid-ask spread while $Q_{-1}$ is empty at the moment of this insertion. 
\item A cancellation of the last limit order at one of the best offer queues.
\item A market order that consumes the last limit order at one of the best offer queues.
\end{itemize}

When $p_{ref}$ changes, the value of $q_{i}$ switches immediately to the value of one of its neighbors (right if $p_{ref}$ increases, left if it decreases). Thus, $q_{\pm 1}$ becomes zero when $p_{ref}$ decreases/increases. Recall that we keep records of the LOB up to the third limit. Consequently, the value for $q_{\pm 3}$ when $p_{ref}$ increases/decreases is drawn from its invariant measure. Note that the queue switching process must be handled very carefully: the average event sizes are not the same for different queues. So, when $q_i$ becomes $q_j$, its new value should be re-normalized by the ratio between the two average event sizes at $Q_i$ and $Q_j$.\\

To possibly incorporate external information, we moreover assume that with probability $\theta^{reinit}$, the LOB state is redrawn from its invariant distribution around the new reference price when $p_{ref}$ changes. The parameter $\theta^{reinit}$ can be understood as the percentage of price changes due to exogenous information. In this case, we consider that market participants readjust very quickly their order flows around the new reference price, as if a new state of the LOB was drawn from its invariant distribution. A similar approach has been used in \citet*{Cont2010} in a model for best bid and best ask queues, in which $\theta^{reinit}$ is set to 1. Under these assumptions, the market dynamics is now modeled by a $(2K+1)$-dimensional Markov process: $\tilde{X}(t):=(X(t), p_{ref}(t))$, in the countable state space $\tilde{\Omega} = \mathbb{N}^{2K} \times \delta\mathbb{N}$, where $X(t) = (q_{-K}(t),...,q_{-1}(t),q_{1}(t),...,q_{K}(t))$ represents the available volumes at different limits.\\

In the sequel, Model I is used to describe the LOB dynamics during periods when $p_{ref}$ is constant (very similar results are obtained in simulations using Model II$^a$ or II$^b$). The $p_{ref}$ change probability $\theta$ and the LOB reinitialization probability $\theta^{reinit}$ are calibrated using the 10 minutes standard deviation of the returns of $p_{mid}$ (the volatility) and the mean reversion ratio $\eta$ introduced in \citet*{robert2011new}, defined by 
$$\eta = \frac{N_c}{2N_a},$$ where $N_c$ is the number of continuations of the estimated $p_{ref}$ on the interval of interest (that is the number of consecutive moves in the same direction) and $N_a$ is the number of alternations (that is the number of consecutive moves in opposite directions)\footnote{Note that here we compute the mean reversion ratio of $p_{ref}$ while the transaction price is usually considered.}. Indeed, the microstructure of large tick assets is well summarized by the parameter $\eta$, see \citet*{robert2011new} and \citet*{dayri2012large} and the volatility is of course one of the most important low frequency statistics. In Figure \ref{fig21}, we show the surfaces of the 10 min volatility and $\eta$ for different values of $\theta$ and $\theta^{reinit}$.\\

\subsubsection{About the maximal mechanical volatility}

\begin{figure}
\centering
\includegraphics[scale=0.65,angle=-90]{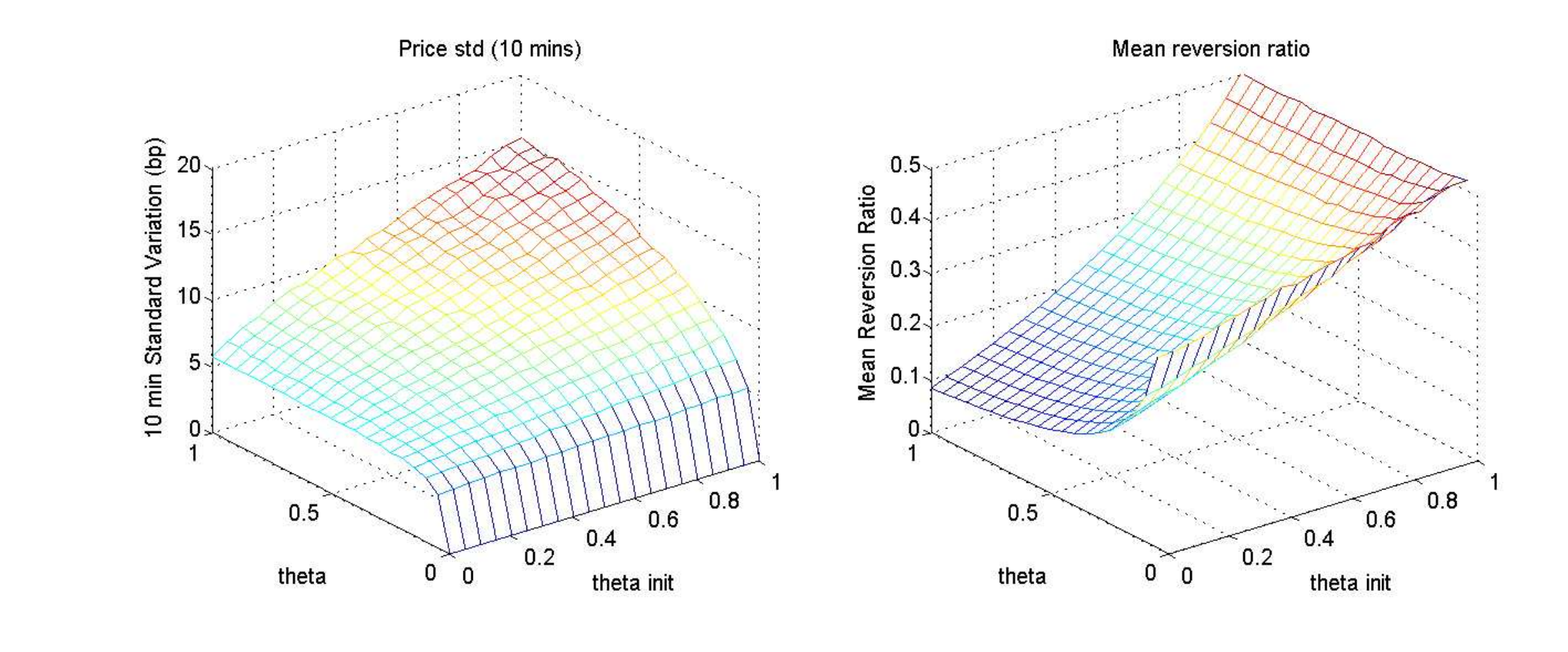}
\caption{10 min volatility and mean reversion ratio, France Telecom} \label{fig21}
\end{figure}

Let us comment now the particular case where we take $\theta^{reinit}=0$. In such situation, Model III becomes a ``purely order book driven model" since the price fluctuations are completely generated by the LOB dynamics. Our simulations show that under this setting, the maximal attainable volatility level (when $\theta = 1$), which we call maximal mechanical volatility, is much lower than the empirical volatility (5 bps compared with 14 bps for the stock France Telecom). This suggests that endogenous LOB dynamics alone may not be enough for reproducing the market volatility. A closer look at these results shows that the model approximates actually quite well the average frequency of price changes, and that the small value of the mechanical volatility is mainly due to the strong mean reverting behavior of the price in this purely order book driven model. This is because of the often reversed bid-ask imbalance immediately after a change of $p_{ref}$. In Figure \ref{fig21}, we can see that the mean reversion ratio $\eta$ is equal to $0.08$ when $\theta = 1, \theta^{reinit} = 0$, which is much smaller than the empirical ratio $0.39$. 

\subsection{Example of application: Order placement analysis}\label{sec:appexp}

We now show how the queue-reactive model can be used in the context of optimal trading. In the general framework of optimal execution, the trading horizon is split into small slices (typically 5-10 minutes) and an algorithm of execution determines the volume to be executed in each slice. This problem, often called ``order scheduling problem'', has been widely studied in the literature, see \citet*{Bertsimas1999, Almgren2000, bouchard2011optimal} for representative examples on this topic. In practice, another optimization issue, the ``order placement problem'', arises naturally once the order scheduling problem has been solved: how should the algorithm place orders to execute the target volume?\\

This second optimization problem can be seen as the microstructural version of the first one. However, it is much more difficult to solve. Indeed, the price dynamics can no longer be approximated by a Brownian motion at these (ultra) high frequency scales. Moreover, the queue priority plays an important role as well as other microstructural features of the asset such as the tick size, the state of the LOB and the trading speed. This in particular implies that execution strategies based on limit or market orders can lead to very different outcomes. Some papers investigate the consequences of using different types of order, see \citet*{harris1996market}, while others aim at finding the best position to place limit orders, see \citet*{laruelle2013optimal}. However, in practice, order placement tactics are usually more complex than the ones considered in the academic literature. For example, traders can hide their trading intentions by splitting furthermore the target volume within each slice. Also, they may start passively, sending limit orders, and then switch to market orders when some market conditions change or a stopping time criteria is met. Very few quantitative tools are available for the analysis of sophisticated tactics and one often needs to rely on so-called market replayers, in which the number of simulations is limited by that of the available trading days in the historical data. Moreover, the market impact, that is the average price drift due to our own trading between the beginning of the execution and a later time, is often neglected. In contrast, our framework is unlimited in number of simulations and is both relevant and easy to use in order to study  market impact profiles and execution costs of complex placement tactics.\\

We write $n_{\rm total}$ for the total quantity to execute and $M$ for the number of slices. An order scheduling strategy gives the target quantity to be executed in each slice, denoted by $n_{i}$ ($n_i \geq 0$ and $\sum_{i=1}^M n_i = n_{\rm total}$). An order placement tactic can be seen as a predefined procedure of order management, ensuring the execution of the target quantity within the slice. Here, as illustration examples, we present two simple tactics, denoted by {\bf T1} and {\bf T2}. In the $i$-th slice, both tactics post a limit order of size $n_i$ at the best offer queue at the beginning of the period, and send a market order with all the remaining quantity to complete the execution of the target volume at the end time of the slice. In between: 

\begin{itemize}
\item {\bf T1  (Fire and forget):} When $p_{mid}$ changes, cancel the limit order and send a market order at the opposite side with all the remaining volume if any.
\item{\bf T2 (Pegging to the best):} When the best offer price changes or our order is the only remaining order at the best offer limit, cancel the order and repost all the remaining volume at the newly revealed best offer queue.
\end{itemize}

Since an order placement tactic is often specifically designed for a given order scheduling strategy, comparisons between two tactics should take into account the associated scheduling strategy, together with the target benchmark\footnote{In execution services, the client often wants the execution algorithm to target some specific price (the arrival price, the average market price during a predefined period,...). The quality of the execution is then assessed on the basis of the difference between the realized execution price and this target benchmark price.}. Other parameters can also have influence when comparing two tactics, such as the total quantity to execute $n_{\rm total}$ and the number of slices $M$. To simplify our analysis, we simulate a buy order of size $n_{\rm total} = 60 \mbox{ AES}_1$, with $M = 20$ and the duration of each slice is fixed to 10 minutes (a total trading period of 3h20). We focus on two benchmarks: the VWAP on the total period (volume weighted average transaction price) and the arrival price $S_{0}$ (the midprice when the execution algorithm starts). Moreover, two types of order scheduling strategies, denoted by {\bf S1} and {\bf S2}, are considered to partly reflect the diversity of optimal trading schemes:

\begin{itemize}
\item {\bf S1:} A linear scheduling ($n_i = {n_{\rm total}}/{M}$), used for the $\mathrm{VWAP}$ benchmark.
\item {\bf S2:} An exponential scheduling $n_i = n_{\rm total}(e^{-(i-1)/4}-e^{-i/4})$, used for the benchmark $S_0$.
\end{itemize}

Finally, note that Assumptions \ref{assump4} and \ref{assump5} are in force for the order cancellation processes.

\subsubsection{Tactic performance analysis}

The performance of an execution algorithm is often measured by its slippage, defined (for a buy order) by 
$$\text{Slippage} = \frac{P_{benchmark}-P_{exec}}{P_{benchmark}}.$$ To understand the effects of the order placement tactic on the execution's slippage, we define the theoretical scheduling slippage by: 

\begin{eqnarray}
P_{exec}^{theo} &=& \sum_{i=1}^M n_i \mathrm{VWAP}^i \nonumber \\
\text{Slippage}^{theo} &=& \frac{P_{benchmark}-P_{exec}^{theo}}{P_{benchmark}}, \nonumber
\end{eqnarray}

where $\mathrm{VWAP}^i$ denotes the volume weighted average transaction price of the $i-$th slice. Indeed, $\mathrm{VWAP}^i$ is often considered as a simple proxy for the execution price in the slice when one focuses on the scheduling algorithm. Hence, $\text{Slippage}^{theo}$ essentially measures the quality of the scheduling strategy and neglects the randomness in execution prices due to the order placement tactic. Note that here a market impact component is included in the computation of the theoretical scheduling slippage. This is because the value of $\mathrm{VWAP}^i$ in each slice is obviously impacted by our execution.\\

\begin{figure}
\centering
\includegraphics[scale=0.65,angle=-90]{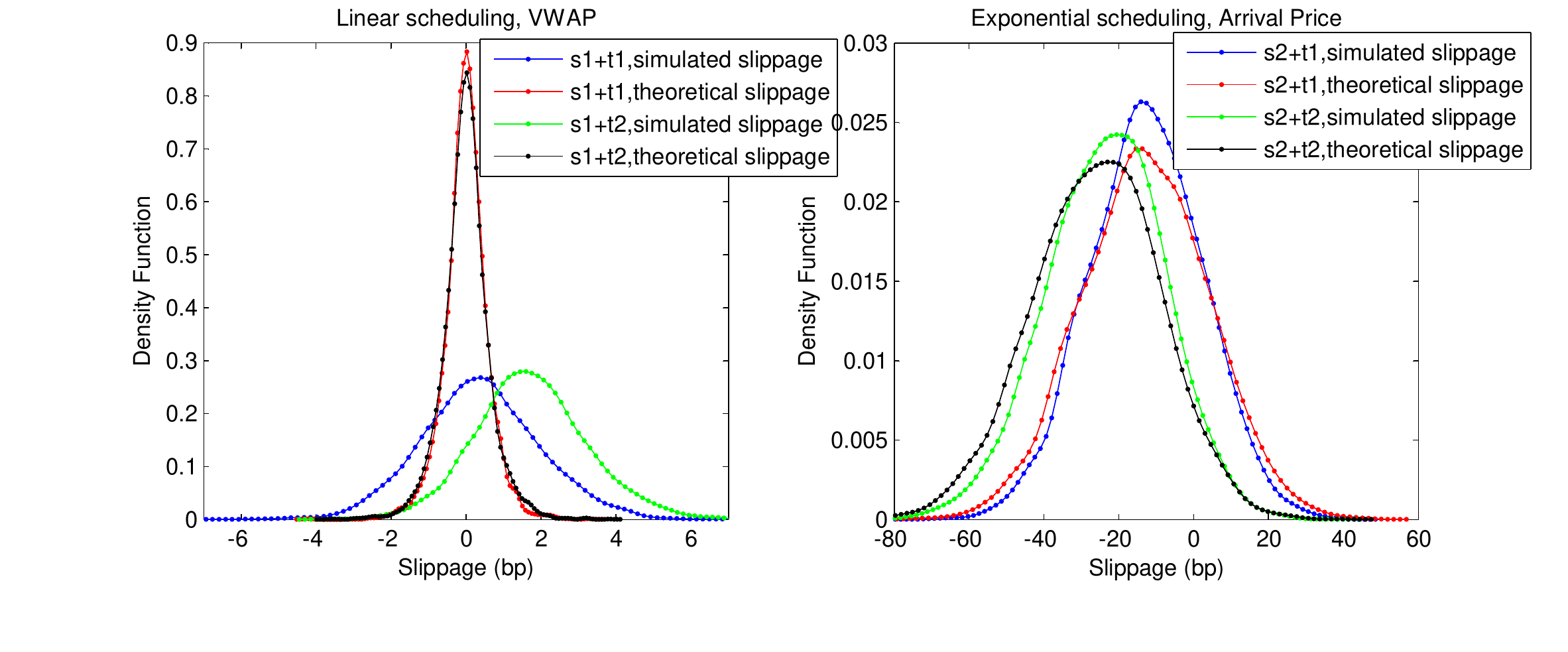}
\caption{Simulation results for the tactics} \label{fig24}

\end{figure}

We launched 2000 simulations for each couple of ({\bf S1/S2}, {\bf T1/T2}). The intensity functions estimated for the stock France Telecom are used in these simulations, as well as the two parameters $\theta=0.7$ and $\theta^{reinit}=0.85$ calibrated in Section \ref{sec:queuereactive}. Furthermore, we use a standard kernel smoothing method when estimating the probability density functions of $\text{Slippage}^{theo}$ and $\text{Slippage}$. The results are shown in Figure \ref{fig24}.\\

Figure \ref{fig24} suggests that the slippage distributions of the same scheduling strategy using two different tactics can be very different: {\bf T2} (``Pegging to the best'') performs better than {\bf T1} (``Fire and forget'') when being coupled with a linear scheduling strategy with VWAP benchmark, while {\bf T1} slightly outperforms {\bf T2} when an exponential scheduling strategy with arrival price benchmark is considered. In our setting, the limit orders change the queue sizes and therefore modify the behaviors of the order flows. Consequently they generate market impact. By constantly following the best offer queue until the total volume is filled, {\bf T2} achieves on average a higher passive execution rate (defined as the volume passively executed\footnote{A buy execution is said to be passive if it occurs at the bid side of the LOB, aggressive if it occurs at the ask side of the LOB.} divided by the total executed volume). Thus, in each slice, it often obtains a better price than that of a more market orders based tactic. However, at the same time, it creates a larger impact than {\bf T1} since the order stays longer in the queues. This explains why the theoretical scheduling slippage of {\bf T2} is worse than that of {\bf T1} for an execution with arrival price benchmark using an exponential scheduling strategy.

\subsubsection{Market impact profiles}\label{sec:marketimpact}

\begin{figure}
\centering
\includegraphics[scale=0.65,angle=-90]{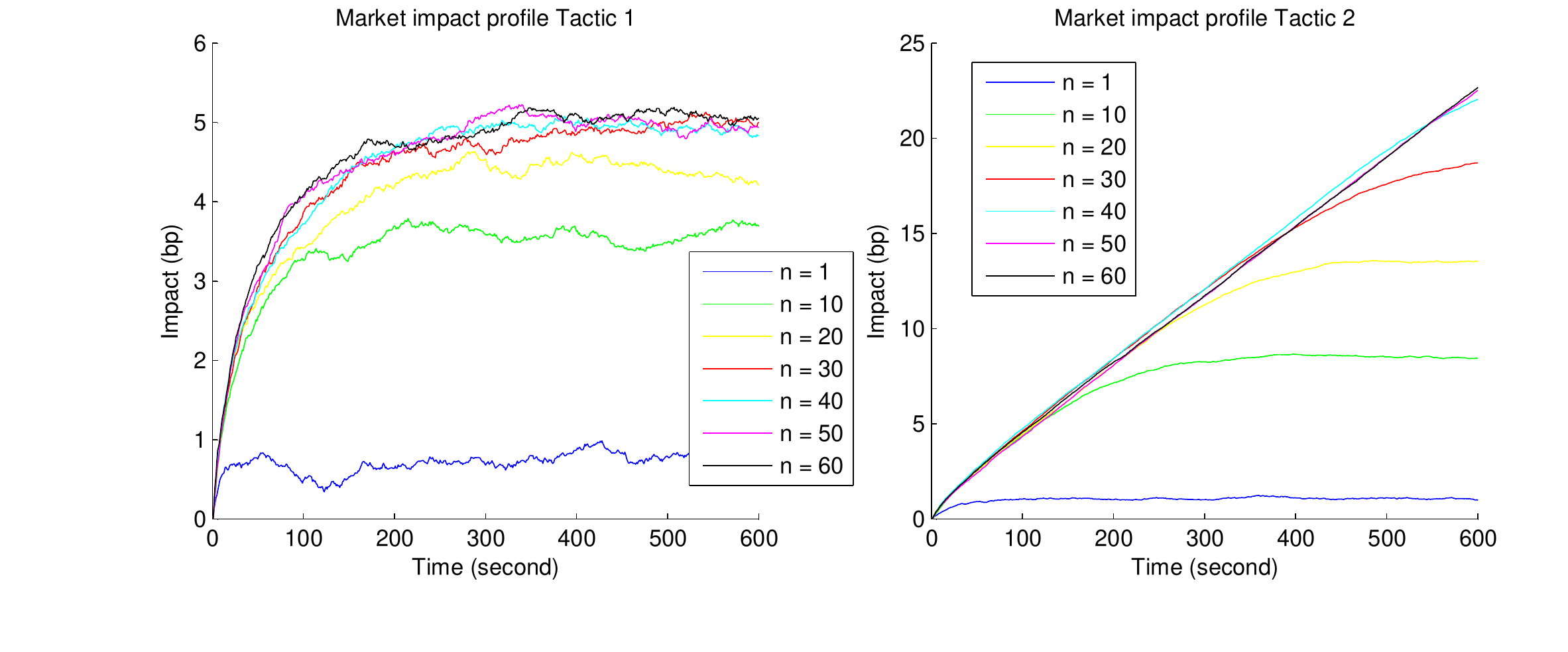}
\caption{Market impact profiles} \label{impactprofile}
\end{figure}

We now study the market impact profiles of these two tactics. Recall that an order placement tactic has two parameters: the slice duration $T$ and the quantity to execute $n$. In the following experiments, $T$ is set to 10 minutes, and the value of $n$ varies from 1 to 60 $\mbox{AES}_1$. We denote by $\mbox{MI}^i(t,n)$ the market impact at time $t$ of Tactic $i$ with target quantity $n$, defined by: $\mbox{MI}^i(t,n) = \mbox{E}[\frac{S_t-S_0}{S_0}]$, with $S_t$ the midprice at time $t$. We launched 2000 simulations for each value of $n,t$ in the ranges $1$-$60$ $\mbox{AES}_1$ and $1$-$600$ seconds. Impact profiles are given in Figure \ref{impactprofile}.\\

In agreement with the celebrated ``square-root law'', see \citet*{gatheral2010no,toth2011anomalous,farmer2013efficiency}, the market impact curves are concave both in time and volume. One can also see that the impact of {\bf T1}  is quite instantaneous and depends essentially on the target quantity $n$, while the impact of {\bf T2} is a progressive process, depending both on the target quantity $n$ and the time $t$. Remark that {\bf T2} seems suitable when dealing with small orders since its market impact is small and it has a higher passive execution rate than {\bf T1}. If one needs to trade larger orders, {\bf T1} becomes probably more relevant since the cost of market impact is likely to outweigh the benefit from passive execution of {\bf T2}. Finally, note that in our Markovian framework, no significant price relaxation (that is the fact that on average, after the completion of the execution of a buy order, the price may drop to a lower level than the one reached at the end of the execution) can be observed.
\section{Conclusion and perspectives}\label{sec:conclusion}

In this work, we have modeled market participants intelligence through their average behaviors towards various states of the LOB. This enabled us to analyze the different order flows and to design a suitable market simulator for practitioners, allowing notably to investigate the transaction costs of complex trading strategies. To our knowledge, our model is the first one where such pre-trade cost analysis is possible in a simple and efficient way.\\

Another important public information, the historical order flow, is not considered in this approach. Market order flows have been shown to be autocorrelated in several empirical studies, see for example \citet*{toth2011order}. Thus, adding such feature in our framework would probably be relevant. Another possible direction for future research would be to explain the shape of the estimated intensity functions in a more sophisticated way. For example, it would be interesting to design some agent based model where these repetitive patterns of the LOB dynamics would be reproduced, providing an even better understanding of the nature of these intensity curves.

\newpage
\section{Appendix}

\subsection{Proof of Theorem \ref{Theo1}}

\begin{proof}
For some $z>1$, set $$V(q)=\sum_{j=-K,j\neq 0}^K{z^{|q_j-C_{bound}|_{+}}}.$$ For any $q \in \Omega$, we have: 
\begin{eqnarray}
\mathcal{Q}V(q) &=& \sum_{p \neq q} \mathcal{Q}_{q,p}[V(p)-V(q)]\nonumber \\
&=& \sum_{i=-K,i\neq 0}^K [f_i(q)(z^{|q_i+1-C_{bound}|_+}- z^{|q_i-C_{bound}|_+})+ g_i(q)(z^{|q_i-1-C_{bound}|_+}-z^{|q_i-C_{bound}|_+})] \nonumber \\
&=& \sum_{i=-K,i\neq 0}^K [f_i(q)1_{q_i\geq C_{bound}}z^{|q_i-C_{bound}|_+}(z-1) + g_i(q)1_{q_i \geq C_{bound}+1}z^{|q_i-C_{bound}|_+}(\frac{1}{z}-1)] \nonumber \\
&=& (z-1) \sum_{i=-K,i\neq 0}^K[f_i(q)1_{q_i \geq C_{bound}}-\frac{g_i(q)1_{q_i \geq C_{bound}+1}}{z}]z^{|q_i-C_{bound}|_+} \nonumber \\
&=& (z-1) \sum_{i:q_i=C_{bound}}f_i(q)+(z-1)\sum_{i:q_i>C_{bound}}[f_i(q)-\frac{g_i(q)}{z}]z^{q_i-C_{bound}}. \label{eqa1}
\end{eqnarray}
Under Assumption \ref{assump1} and \ref{assump2}, we can find a $z$ sufficiently close to 1 such that, if $q_i>C_{bound}$,
\begin{equation*}
f_i(q)-\frac{g_i(q)}{z} < z^{-1}(-r+H(z-1)) = -r'< 0.
\end{equation*}
So, from Equation \eqref{eqa1}, we have 
\begin{eqnarray*}
\mathcal{Q}V(q) &\leq& (z-1)H - (z-1)\sum_{i:q_i>C_{bound}}r'z^{q_i-C_{bound}} \nonumber \\
&\leq& -(z-1)r'\sum_i z^{|q_i-C_{bound}|_+} + (z-1)H+2(z-1)r'K \nonumber \\
&\leq& -(z-1)r'V(q) + (z-1)[H+2r'Kz]. \label{eqa2}
\end{eqnarray*}
Thus $X(t)$ is V-uniformly ergodic. Then using Theorem 4.2 in \citet*{meyn1993stability}, $X(t)$ is Harris positive recurrent and has a finite invariant measure. Furthermore, by Theorem 3.6.2 in \citet*{norris1998markov}, the process $X(t)$ converges to its equilibrium and is therefore ergodic. 
\end{proof}

\subsection{Computation of confidence intervals}

When the queues are independent, by the central limit theorem, we have, with asymptotic probability 95\% (we note $\hat{p_i}^L(n) = \frac{\# \{\etype(\omega)\in\mathcal{E}^+, q_i(\omega) = n\}}{\# \{q_i(\omega) = n\}}$):

\begin{eqnarray}
\Lambda_i(n) &\in& [\hat{\Lambda}_i(n) - \frac{1.96\hat{\Lambda}_i(n)}{\sqrt{\#\{q_i(\omega) = n\}}}, \hat{\Lambda}_i(n) + \frac{1.96\hat{\Lambda}_i(n)}{\sqrt{\#\{q_i(\omega) = n\}}}] \nonumber \\ \nonumber
\frac{\lambda_i^L(n)}{\Lambda_i(n)} & \in & [\hat{p_i}^L(n) - \frac{1.96\sqrt{\hat{p_i}^L(n)(1-\hat{p_i}^L(n))}}{\sqrt{\#\{q_i(\omega) = n\}}}, \hat{p_i}^L(n) + \frac{1.96\sqrt{\hat{p_i}^L(n)(1-\hat{p_i}^L(n))}}{\sqrt{\#\{q_i(\omega) = n\}}}].
\end{eqnarray}

So, at least with probability 90\%:
\begin{eqnarray}
\lambda_i^L(n) &\in& [(\hat{\Lambda}_i(n) - \frac{1.96\hat{\Lambda}_i(n)}{\sqrt{\#\{q_i(\omega) = n\}}}) (\hat{p_i}^L(n) - \frac{1.96\sqrt{\hat{p_i}^L(n)(1-\hat{p_i}^L(n))}}{\sqrt{\#\{q_i(\omega) = n\}}}) \nonumber \\ & & (\hat{\Lambda}_i(n) + \frac{1.96\hat{\Lambda}_i(n)}{\sqrt{\#\{q_i(\omega) = n\}}}) (\hat{p_i}^L(n) + \frac{1.96\sqrt{\hat{p_i}^L(n)(1-\hat{p_i}^L(n))}}{\sqrt{\#\{q_i(\omega) = n\}}})].   \nonumber
\end{eqnarray}

Similar results can be computed for $\lambda_i^C$ and $\lambda_i^M$. The method used to compute confidence intervals of Model II$^a$ is quite similar. Confidence intervals are more difficult to compute in Model II$^b$, and we use approximations by neglecting the possible intersections between the two sets: $\{q_1(\omega) = n,\Size_{m,l}(q_{-1}(\omega))\in s)\}$ and $\{q_{-1}(\omega) = n,\Size_{m,l}(q_{1}(\omega))\in s)\}$.

\subsection{Quasi birth and death process}
\begin{definition}
(Quasi birth and death process, from \citet*{latouche1999introduction}): A quasi birth and death (QBD) process is a bivariate Markov process with countable state space $S = $ \{$(i,j):i\geq 0, j = 0,1,...,m$\} where the first element $i$ is called the level of the process, and the second element $j$ is called the phase of the process. The parameter $m$ can be either finite or infinite. The process is restricted in level jumps only to its nearest neighbors, meaning that the probability of jumping from level $i$ directly to level $l, l \geq i + 2$ or $l \leq i-2$ is equal to zero.
\end{definition}
We can easily see that the Markov process $(q_1,q_2)$ in Model II$^a$ is indeed a QBD process with countable phases.
Its infinitesimal generator matrix is of the following form:
\[ Q = \left[ \begin{array}{ccccc}
A_1^{(0)} & A_0^{(0)} & 0 & 0 & ... \\
A_2^{(1)} & A_1^{(1)} & A_0^{(1)} & 0 & ... \\
0 & A_2^{(2)} & A_1^{(2)} & A_0^{(2)} & ... \\
... & ... & ... & ... & ... \\
 \end{array} \right],\]
where the matrix $A_0^{(\ell)}$ encodes transitions from level $q_1=\ell$ to level $q_1=\ell+1$, matrix $A_2^{(\ell)}$ encodes transitions from level $q_1=\ell$ to level $q_1=\ell-1$, and matrix $A_1^{(\ell)}$ encodes transitions within level $q_1=\ell$. More specifically, the element ($i,j$) of $A_0^{(\ell)}$ is the transition rate from state $(q_1=\ell,q_2=i)$ to state $(q_1=\ell+1,q_2=j)$, the element ($i,j$) of $A_2^{(\ell)}$ is the transition rate from state $(q_1=\ell,q_2=i)$ to state $(q_1=\ell-1,q_2=j)$, and the element ($i,j$) of $A_1^{(\ell)}$ is the transition rate from state $(q_1=\ell,q_2=i)$ to state $(q_1=\ell,q_2=j)$. \\

\noindent We write the intensity functions at $Q_2$ when $q_1 = 0$ with a $\tilde{}$. For matrix $A_i^{(\ell)}$, $i=0,1,2$, we have: 
\begin{eqnarray}
A_0^{(k)} &=& \lambda_1^L(k) I \nonumber,\\
A_2^{(k)} &=& (\lambda_1^C(k) + \lambda_{buy}^M(k)) I, \nonumber
\end{eqnarray}
\[ A_1^{(0)} = \left[ \begin{array}{cccc}
-\lambda_1^L(0)-\tilde{\lambda}_2^L(0) & \tilde{\lambda}_2^L(0) & 0 & ...  \\
\tilde{\lambda}_2^C(1) + \tilde{\lambda}_{buy}^M(1) & - \lambda_1^L(0) - \tilde{\lambda}_2^L(1) - \tilde{\lambda}_2^C(1) - \tilde{\lambda}_{buy}^M(1) & \tilde{\lambda}_2^L(1) & ... \\
0 & \tilde{\lambda}_2^C(2) + \tilde{\lambda}_{buy}^M(2) & - \lambda_1^L(0) - \tilde{\lambda}_2^L(2) - \tilde{\lambda}_2^C(2) - \tilde{\lambda}_{buy}^M(2) & ... \\
... & ... & ... & ...  \\
 \end{array} \right],\]
and for $k \geq 1$:
\[ A_1^{(k)} = \left[ \begin{array}{cccc}
-\lambda_1^C(k) -\lambda_{buy}^M(k) -\lambda_1^L(k)-\lambda_2^L(0) & \lambda_2^L(0) & 0 & ...  \\
\lambda_2^C(1) & -\lambda_1^C(k) -\lambda_{buy}^M(k) -\lambda_1^L(k)-\lambda_2^L(1)-\lambda_2^C(1) & \lambda_2^L(1) & ...  \\
... & ... & ... & ...  \\
 \end{array} \right].\]

We define $\pi_{i,j} = \mathbb{P}[q_1 = i, q_2 = j]$ the stationary distribution of this QBD process, and:
\begin{eqnarray}
\pi_n &=& [\pi_{n,0}, \pi_{n,1}, ... ] \nonumber\\
\pi &=& [\pi_0, \pi_1, ...]. \nonumber
\end{eqnarray}
We shall have:
\begin{eqnarray*}
\pi Q &=& 0 \nonumber \\
\pi 1 &=& 1.
\end{eqnarray*}

The dynamics of the two queues system $(q_1,q_2)$ is level dependent, meaning that its transition kernel depends on the value of $q_1$. This makes the computation or approximation of its asymptotic behavior quite difficult. Thus we consider an additional assumption in order to turn $(q_1,q_2)$ into a so-called level independent QBD process. This is particularly interesting since it enables us to easily express the invariant measure in a matrix geometric form and to compute it numerically. The level independence property is defined by the fact that for $i\geq 1$, $A_0^{(i)}$, $A_1^{(i)}$ and $A_2^{(i)}$ do not depend on $i$ and  $A_0^{(0)}=A_0^{(i)}$, see \citet*{latouche1999introduction}. Under the following assumption, this property is satisfied by $(q_1,q_2)$ in Model II$^a$. 

\begin{assumption} \label{assump3}
(Independent Poisson flows at first limits) There are two positive constants $\lambda_1$ and $\mu_1$, with $\lambda_1 < \mu_1$, such that for $k \geq 1$: 

\begin{eqnarray*}
\lambda_1^C(k) + \lambda_{buy}^M(k) &=& \mu_1 \nonumber \\
\lambda_1^L(k) &=& \lambda_1 \nonumber \\
\lambda_1^L(0) &=& \lambda_1.
\end{eqnarray*}
\end{assumption}

In practice, $\lambda_1$ and $\mu_1$ are taken as the average values of the estimated intensity functions at first limits. Under this assumption, a quite simple numerical computation of the invariant distribution is possible. Generally speaking, QBD processes with finite phase (meaning that the value set of the second dimension, in our case $q_2$, is finite) can be easily treated, see for example \citet*{latouche1999introduction}. In the infinite case, truncation methods must be applied to obtain approximate results. Thanks to the special structure of the generator in our model, one simple truncation method, called ``first column augmentation by block'', can be applied. Details of this truncation method can be found in \citet*{bean2010approximations}. The mathlab toolbox SMCSolver, see \citet*{bini2006structured}, is used to compute the invariant measure. 

\subsection{Alcatel-Lucent}
Results for the stock Alcatel-Lucent are presented in the following figures (Figure \ref{alc1} to Figure \ref{alc2}).
\begin{figure}
\centering
\includegraphics[scale=0.65,angle =-90]{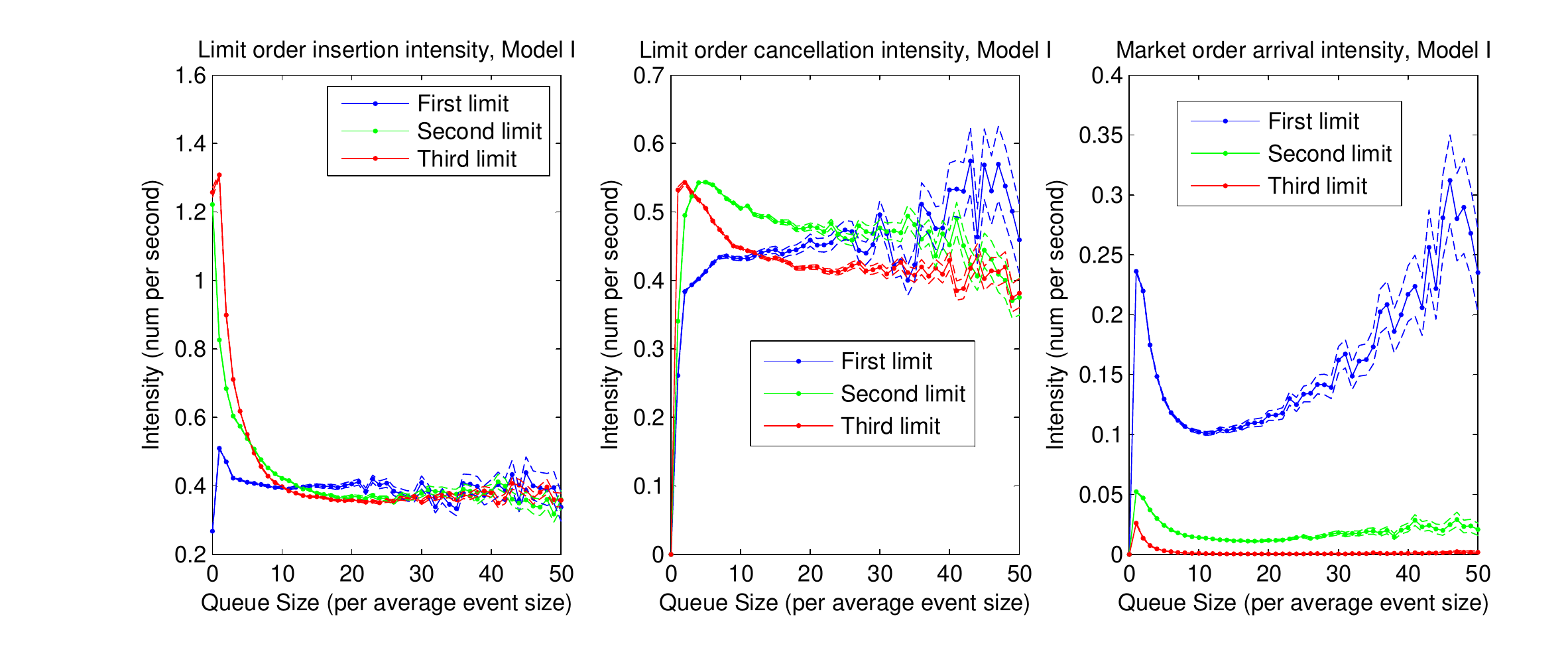}
\caption{Intensities at $Q_{\pm 1,2,3}$, Alcatel Lucent}\label{alc1}
\end{figure}
\begin{figure}
\centering
\includegraphics[scale=0.65,angle =-90]{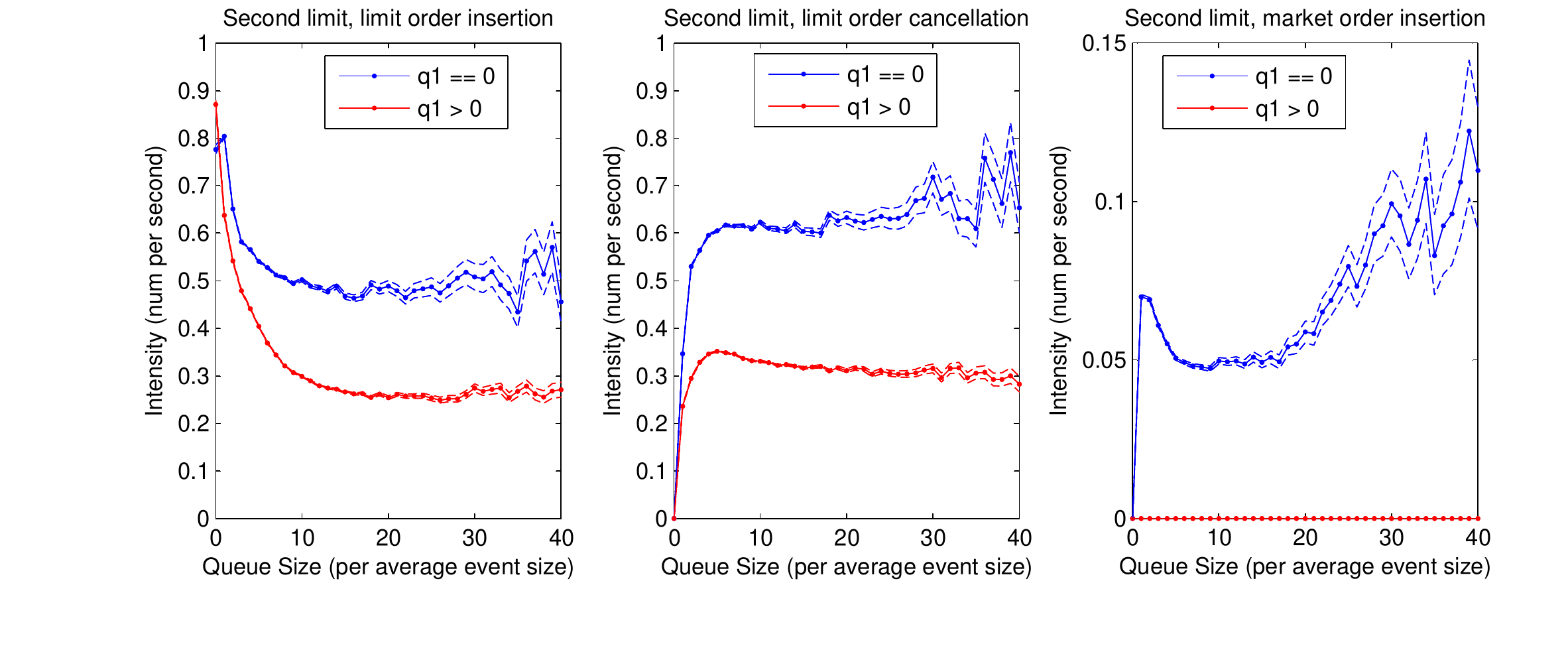}
\caption{Intensities at $Q_2$ as functions of $1_{q_1>0}$ and $q_2$, Alcatel Lucent}
\end{figure}

\begin{figure}
\centering
\includegraphics[scale=0.65,angle =-90]{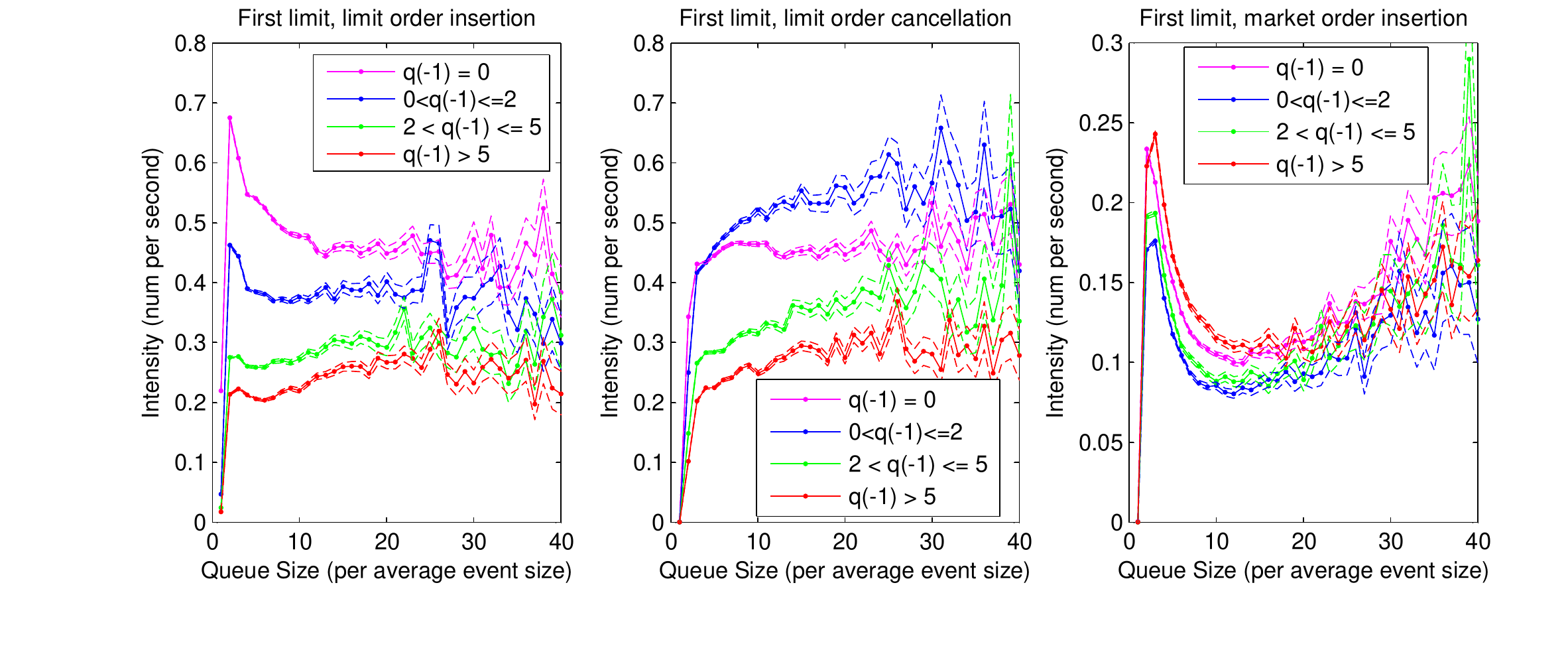}
\caption{Intensities at $Q_1$ as functions of $\Size_{m,l}({q_{-1}})$ and $q_1$, Alcatel Lucent}
\end{figure}

\begin{figure}
\centering
\includegraphics[scale=0.65,angle =-90]{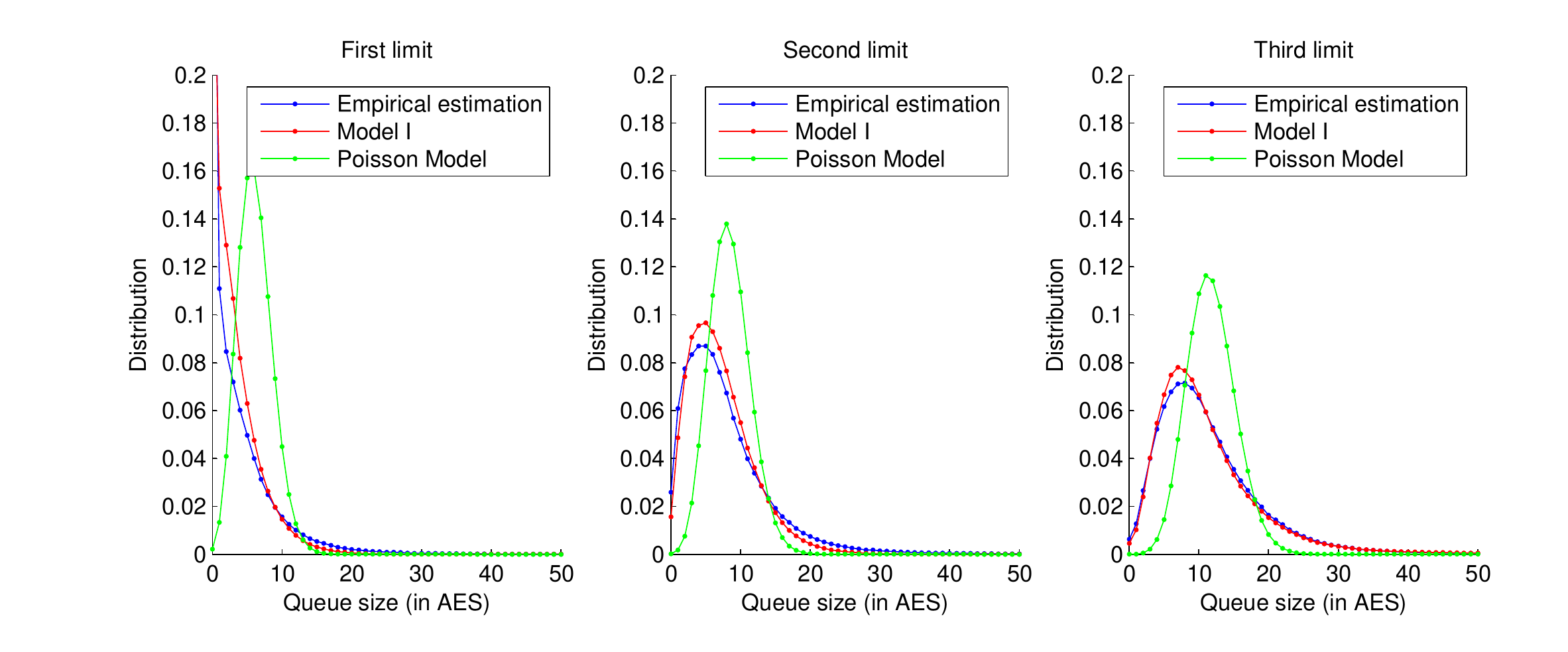}
\caption{Queue distribution, Alcatel Lucent}
\end{figure}

\begin{figure}
\centering
\includegraphics[scale=0.65,angle =-90]{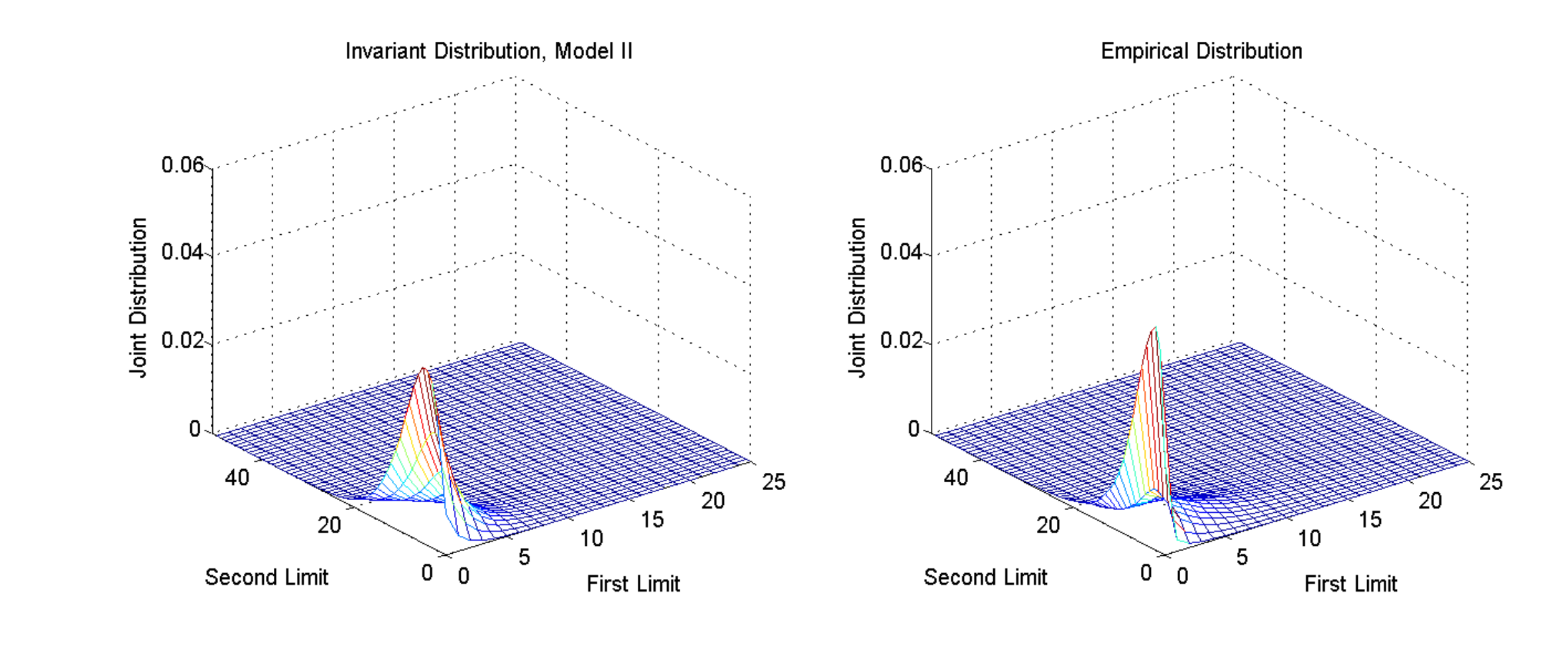}
\caption{Model II$^a$: joint distribution of $q_1,q_2$, Alcatel Lucent}
\end{figure}

\begin{figure}
\centering
\includegraphics[scale=0.65,angle =-90]{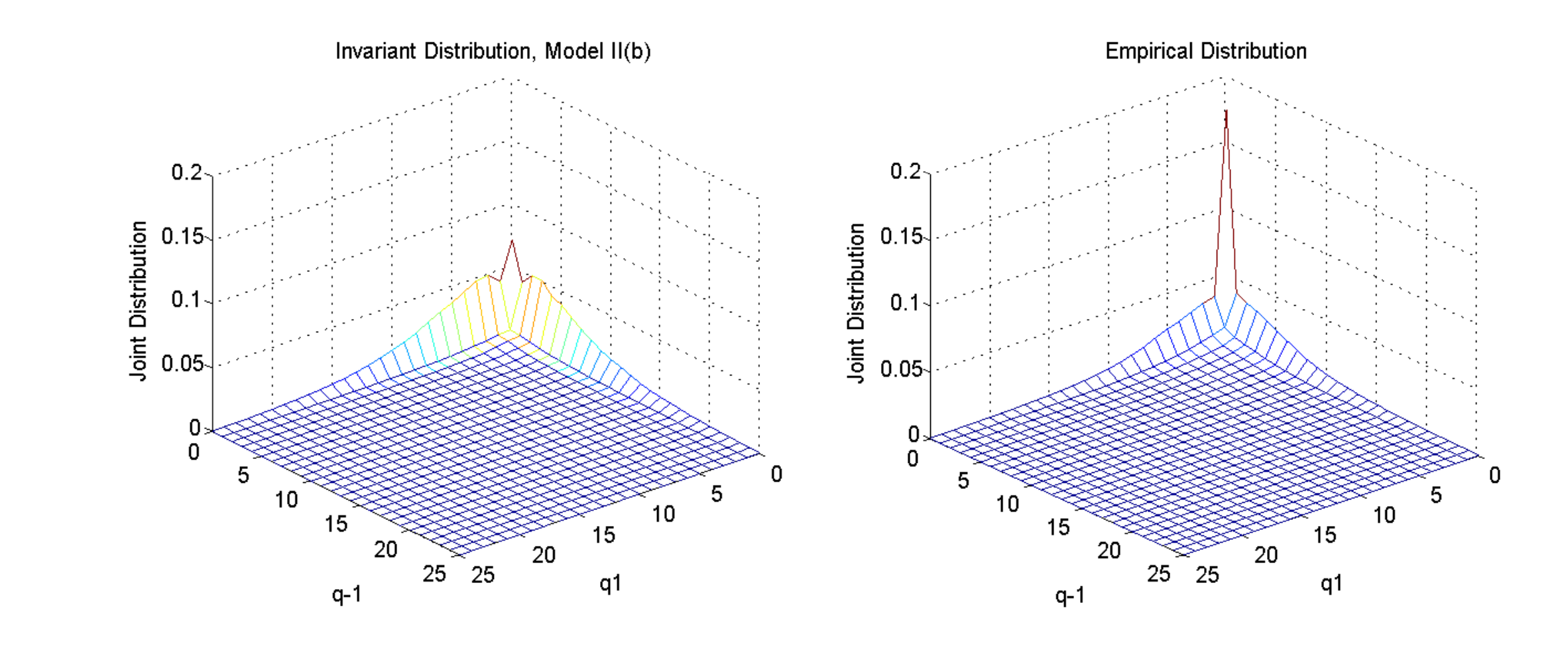}
\caption{Model II$^b$: joint distribution of $q_{-1},q_1$, Alcatel Lucent}\label{alc2}
\end{figure}

\subsection{AES}\label{sec:aes}

$\mbox{AES}_i$ is defined as the average size of all events (including limit order insertion, cancellation and trades) at $Q_i$, while ATS computes only the average size of all trade events. In Table \ref{aesandats} we show the estimated values of AES at different distances to $p_{ref}$ and the estimated value of ATS, for the stocks France Telecom and Alcatel-Lucent. 

\begin{table}[h]\footnotesize
 \label{aesandats}
\begin{center}
\begin{tabular}{| l | l | l | l | l |}
\hline
stock & ATS & $\mbox{AES}_1$ & $\mbox{AES}_2$ & $\mbox{AES}_3$ \\ \hline
France Telecom & 637 & 836 & 1068 & 1069 \\ \hline
Alcatel Lucent & 2340 & 3033 & 3451 & 3528 \\ \hline
\end{tabular}
\caption{AES and ATS (in number of stocks)}
\end{center}
\end{table}

\bibliography{ref}

\end{document}